\documentclass[pra,twocolumn]{revtex4}
\usepackage{amsthm}
\usepackage{amsmath}
\usepackage{latexsym}
\usepackage{amsfonts}
\usepackage{amssymb}
\usepackage{color}
\usepackage{bbm,dsfont}
\usepackage{graphicx}
\usepackage{hyperref}
\usepackage{subfigure}

\newcommand{\id}{\mathbbm{1}} 

\def\id{{\mathbb I}}

\mathchardef\ree="023C \mathchardef\imm="023D  

\def\OO{{\mathcal O}}

\newcommand{\nn}{{\langle N \rangle}}
\newcommand{\nnx}{{\scriptscriptstyle{\langle N \rangle}}}

\begin{document}
\title{
Fisher information from quantum many-particle arrival time measurements}

\author{Jukka Kiukas}
\affiliation{Department of Mathematics, Aberystwyth University, Aberystwyth SY23 3BZ, UK}
\author{Andreas Ruschhaupt}
\affiliation{School of Physics, University College Cork, Cork, Ireland}

\begin{abstract} We formulate a quantum arrival time measurement process for a Bosonic many-particle system, with the aim of extracting statistical information on single-particle properties. The arrival time is based on a dynamical multi-particle absorption model in the Fock space, and we consider systems in coherent and incoherent mixtures of $N$-particle states. We find the resulting probability distributions for arrival time sequences, which we consider as parametric models for the statistical inference of single-particle parameters, and derive a tractable expression for the associated (classical) Fisher information. Subsequently focusing on the concrete case of the momentum parameter of a 1D particle, we consider the idealized limits of a point (Dirac delta) detector and an infinite particle system forming a spatially uniform ``beam''. We observe that even though no information remains in the spatial distribution, the single-particle momentum is indeed identifiable from the arrival time data, even in the limit of ``sparse beams'' of vanishing particle density, where we obtain simple analytical form for the Fisher information, which, interestingly, coincides with the one obtained from a hypothetical time-stationary detection model. Our results contribute to the fundamental understanding of temporal measurement data arising from quantum systems consisting of freely evolving particles.
\end{abstract}

\maketitle

\section{Introduction} 

Since the paper consists of several interlinked stages based on various background literature, we describe both briefly in the present section before proceeding to the detailed content.

\subsection{Context}
While detection times are measured routinely in experiments, the problem of describing \emph{time as an observable} rigorously within the quantum theoretic formalism is far from being straightforward. Indeed, the problem has been studied since the beginning of quantum mechanics \cite{Muga2000, Muga2002, Muga2009}
and continues to attract interest \cite{Kiukas2009, Kiukas2013, Maccone2020, Roncallo2023, Kijowski2023, Page1983, Giovannetti2015,Marletto2017}.

In typical textbooks of basic quantum mechanics, the problem is avoided by only considering time as a parameter of the evolution of a quantum system. However, such texts nevertheless often suggest, based on the formalism of Fourier transforms and the related Heisenberg uncertainty principle, that time and energy are ``conjugated'' in the same way as position and momentum of a free particle. However, this is not true in the usual sense, as there is no selfadjoint operator canonically conjugate to a Hamiltonian with lower bounded spectrum; this result is known as Pauli's theorem \cite{Pauli1958}. Therefore any physically relevant system with a finite ground state energy cannot have a selfadjoint ``time operator''.

Several approaches have been proposed to get around this problem, depending on the type of time considered \cite{Muga2002, Muga2009}. We focus here on \emph{time of arrival}, which has a clear operational interpretation in terms of detection schemes, where e.g. a quantum particle arrives at a detector screen. In this case there are several approaches with varying level of abstraction; the most abstract approach focuses on finding a workaround to the non-existence of a self-adjoint time operator, by relaxing the requirement of selfadjointness, and instead only requires that an arrival time observable is represented by a positive operator valued measure (POVM) covariant for the energy observable of the system. In this setting one still obtains natural time-energy uncertainty relations, and explicit solutions have been constructed, in particular the Aharonov-Bohm time-of-arrival operator \cite{Aharonov1961} and the related Kijowski time-of-arrival distribution \cite{Kijowski1974};
for a review see \cite{Muga2002a}. The Kijowski time-of-arrival distribution can be seen as an instance of a general energy-covariant observable \cite{Werner1986}, and this approach can be further generalised to multi-particle arrival time \cite{Werner1989, Baute2002}, and further to the case of an infinite time-stationary particle ``beam" \cite{Kiukas2013}.

Another approach to arrival times is to consider a microscopic detector model for the quantum time of arrival, for instance an atom-laser model \cite{Damborenea2002, Ruschhaupt2009}.

A third approach, which we follow in this paper, lies between the above two in terms of the level of abstraction; the idea is to use the theory of open quantum systems to build a phenomenological model for the detection process \cite{Allcock1969,Werner1987, Ruschhaupt2004, Ruschhaupt2004a}, which still satisfies e.g. natural energy-time uncertainty relations \cite{Kiukas2012}.
Such a detection process can be described in terms of repeated observations of a quantum system during its evolution; this framework has appeared in different forms in several areas of open quantum systems, as the ``quantum jump’’ or "quantum trajectories" model in quantum optics \cite{Hegerfeldt1992,Dalibard1992,Carmichael1993},
the operational theory of quantum measurements and continual observations \cite{Davies1976, Barchielli1986},
the theory of quantum Markov processes and collision models \cite{Ciccarello2022}, and finitely correlated or matrix product states in a chain of quantum systems \cite{Fannes1992, Verstraete2010}.

We note that there are also alternative approaches to obtain detection times, see for example \cite{Benard1973} for photon detection, but these are less relevant to the context of our work.

\subsection{Aims of the paper}

Here we follow the third approach above to build a phenomenological model of our arrival time detection process. We will consider a many-particle system undergoing absorptive dynamics, where each absorption event corresponds to an arrival of a particle to the detector; the particle is subsequently destroyed and the arrival time is recorded. In this way each detection removes one particle from the system, and we get a record of arrival times as a result; this is conveniently described in a Fock space setting using the canonical annihilation operators \cite{Bratteli1997}. 
We specifically restrict to the case of massive Bosonic particles, and focus in detail an application on particles moving in one dimension.
When one wishes to separate detection from the source, it is natural to describe the state prepared by the source in a Fock space (many particle system), and formulate instead the dynamics of the absorption model at this system. For this we use the master equation introduced in \cite{Alicki2007, Butz2010}, apply the quantum jump method to unravel it, and hence extract the detection time distributions.

There are obviously many ways of composing many-particle states form the single-particle ones. We focus on the cases where the many-particle state is a certain finite-particle, coherent, or incoherent combination of single-particle wave-functions. \emph{Our main aim is, then, to investigate how single-particle properties can be inferred from such many-particle states using arrival time data}, and for that we use the standard framework of statistical estimation, see e.g. \cite{Garthwaite2006}.

Statistical estimation is one of the fundamental ingredients of quantum technology \cite{Paris2009}. The general idea is to infer the unknown quantum state (in a suitable parameterisation) from the statistics of relevant measurements performed on the system, by describing how the theoretical probability distribution (the statistical model) of the measurement outcome depends on the parameter, and then fitting the model to the observed data using suitable estimators. Then the error of any unbiased estimator (quantified by variance) is bounded below by the inverse of the \emph{Fisher information} of the parameter. Therefore, Fisher information is considered as a quantifier of the information on the true value of the parameter obtainable using the measurements in question. Of course, this crucially depends on the statistical model, which will be given by the theory describing the relevant measurements. Here we consider the estimation of parameters which specify the single-particle wavefunction of a many-particle state, using arrival time measurements described above. Up to our knowledge, estimation based on arrival times has not been systematically studied in the literature. It is especially interesting due to its indirect character -- the parameters are inferred through the effect on particle propagation rather than directly measuring the state of the particles. The results provide insight into the use of arrival time statistics in metrology based on particle systems evolving freely in space, in contrast to e.g. optical waveguides \cite{Bratteli1997}.

\subsection{Structure of the paper and outline}

\begin{figure}
\begin{center}
(a)
\includegraphics[width=0.92\linewidth
]{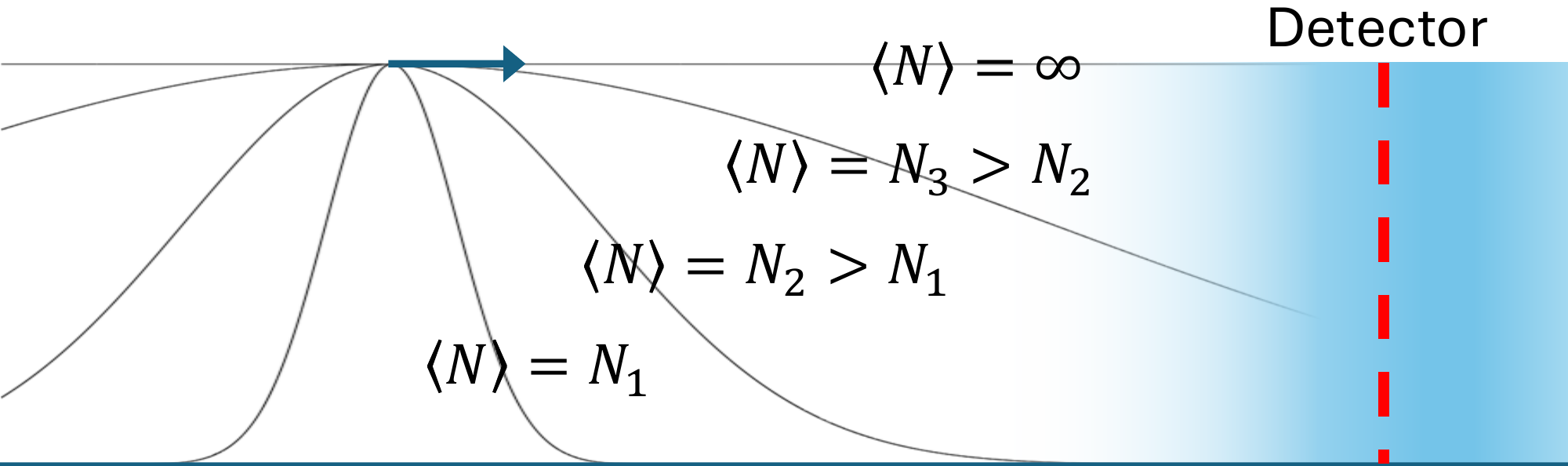}

\

(b) \includegraphics[width=0.92\linewidth
]{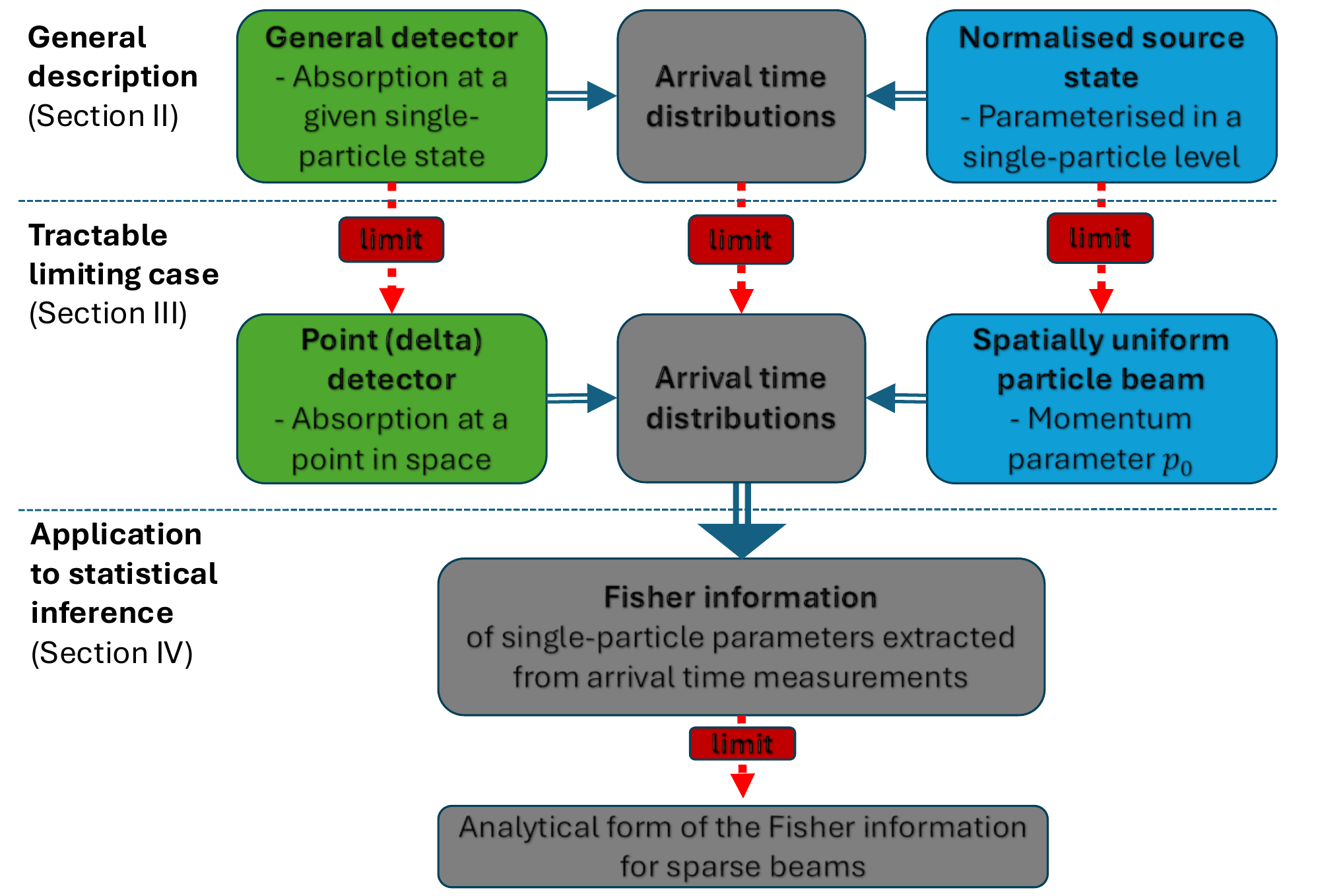}
\end{center}
\caption{\label{fig:context} (a) An illustration of a beam of particles with arrival time detection. The detection takes place in blue region, mathematically described by the loss of the wavefunction normalisation giving rise to detection probabilities, and followed by particle annihilation in the Fock space. In the case of a Dirac delta detector, the annihilation takes place in a single point in space (red line). On the source side, suitably scaled position distributions of different initial single-particle states $\chi$ are shown depending on the mean particle number $\langle N\rangle$, and with the limit $\langle N\rangle=\infty$ corresponding to a spatially uniform beam localised at single momentum $p_0$. (b) A schematic figure showing the structure of the paper: we first present our general detection scheme, then work out the limiting cases illustrated in (a), and finally apply statistical inference to quantify information on $p_0$ obtained from the arrival time data.} 
\vspace{-0.5cm}
\end{figure}

The idea of an absorbing arrival time detector is illustrated in Fig. \ref{fig:context}(a) using our main specific setting of a system of massive free 1D Bosons detected at a single point in the position space. As the mean particle number $\langle N\rangle$ grows the single-particle position distribution widens until it becomes flat at the limit $\langle N\rangle=\infty$, corresponding to a spatially uniform ``beam" of fixed particle density $r_0$, localised in the momentum space at a single value $p_0$, which we wish to estimate from arrival time data. 

Fig. \ref{fig:context}(b) shows the overall structure of the paper, which consists of three stages. As the first step, in Section \ref{sec:detectormodel}, we formulate the general scheme, i.e. we describe rigorously the particle absorption model for suitable initial states in Fock space, and derive the corresponding arrival time distributions. We then focus on the setting of 1D particles, which provides a tractable concrete case where we obtain explicit analytical results illustrating the features of the detection model, including how the information on the relevant parameters depends on the number of detections. 

In the second step (Section \ref{sec:deltabeamlimits}) we first introduce the point detector limit, which is in a sense a canonical choice if we do not assume anything particular about the detection process. Heuristically, this corresponds to a Dirac delta function, and we will describe it rigorously as a suitable limit of normalised states, deriving an analytical formula for the arrival process. After this, we consider the analogous canonical choice for the single-particle state, namely a beam of uniform spatial particle density $r_0$ localised in a single momentum $p_0$;
this setting was already described above.
In order to take the limit, we need to go beyond the Fock space description as shown in Fig. \ref{fig:context}(a), because uniform spatial particle density with definite momentum is only possible with infinite particle number. Here we do not consider the resulting states (which would be non-normal states on a suitable abstract operator algebra \cite{Bratteli1997}), but rather follow a concrete approach by showing that the arrival time distributions converge at the limit, which must be taken so that for large $\langle N\rangle$, the area under the curve in Fig. \ref{fig:context}(a) is approximately equal to $\langle N\rangle/r_0$.

As the third step, in Section \ref{sec:metrology}, we introduce the framework for statistical estimation of single-particle properties using arrival time statistics, derive a tractable formula for the classical Fisher information quantifying the optimal estimation accuracy, and use it for our main setting of spatially uniform beam. Here we have one relevant single-particle parameter, $p_0$, to estimate, and furthermore, we know that the position statistics do not provide any information on $p_0$. Hence, studying how much information on $p_0$ can be extracted from temporal (i.e. arrival time) statistics is especially interesting. Finally, we consider the simplifying limit of ``sparse beams'', where the particle density $r_0$ tends to zero, and derive a simple analytical form for the Fisher information.

Each of the main three sections include a subsection summarising the main results of the corresponding step and presenting some illustrating examples. Finally, we conclude the paper in Section \ref{sec:outlook} with a summary of the main observations and further directions.

\section{Particle detection model}\label{sec:detectormodel}
Here we describe in detail the phenomenological model of the detector. We first review single-particle detection, then set up the relevant master equation in Fock space, and finally derive the arrival time distributions for suitable many-particle states by unravelling the dynamics.

\subsection{Single-particle detection}\label{subsec:sp}

In order to review the basic idea of a phenomenological detector model, we begin with the case where the observed system consists of only one particle \cite{Allcock1969,Werner1987,Ruschhaupt2004,Ruschhaupt2004a, Kiukas2012}. The basic ingredients are a given time-independent Hermitian Hamiltonian $H$ describing a free evolution of a system with Hilbert space $\mathcal H$, a fixed \emph{absorbing state} $\phi\in \mathcal H$ with $\|\phi\|=1$, and a constant $\gamma>0$ describing the detection strength.

The detection event occurs at random time $t$, whose distribution is obtained as follows: we first initialise the particle in a vector state $\chi\in \mathcal H$, which we call the \emph{source state}. We then define the contraction semigroup $U_t = e^{-it H_{\rm eff}/\hbar}$ using the \emph{effective Hamiltonian}
\begin{equation}\label{Heff}
H_{\rm eff} = H -\frac{i\hbar}{2}L^\dagger L = H-i\gamma \frac{\hbar}{2} |\phi\rangle\langle \phi|
\end{equation}
with an imaginary part, which in our case is always rank one as shown. Here the operator $L$ can be used to specify what happens to the system after the detection, but only $L^\dagger L$ contributes to the semigroup. This gives an ``evolution''
\begin{equation}\label{chit}
\chi_t:= U_t\chi,
\end{equation}
where \emph{the loss of normalisation $1-\|\chi_t\|^2$ is interpreted as the probability that the particle is detected before time $t$}. The phenomenological idea is that the non-hermitian part in \eqref{Heff} models the effect of the detector on the particle which dynamically leads to loss of the norm of the wave function. The probability density of the detection time is therefore
\begin{align}
p(t) &= \frac{d}{dt} (1-\|\chi_t\|^2)= -\frac{d}{dt} \langle U_t\chi|U_t\chi\rangle\nonumber\\
&=\frac{i}{\hbar}\langle \chi_t|(H_{\rm eff}-H_{\rm eff}^\dagger)\chi_t\rangle = \gamma |\langle \phi|\chi_t\rangle|^2.\label{singleprob}
\end{align}

Hence, the relevant quantity is $\langle \phi|\chi_t\rangle$; in order to find it we will use the formula
\begin{equation*}
U_t = e^{-itH/\hbar}- \frac {\gamma}{2}\int_{0}^t e^{-i(t-s)H/\hbar}|\phi\rangle\langle \phi|U_s ds,
\end{equation*}
which is easy to verify by checking that it satisfies the defining equation $i \hbar \frac{d}{dt}U_t = H_{\rm eff}U_t$. This implies
\begin{equation}\label{integraleq}
h(t) = h_0(t) - \frac {\gamma}{2} \int_0^t g(t-s) h(s) ds,
\end{equation}
where $h(t) = \langle \phi|\chi_t\rangle$, $h_0(t) = \langle \phi|e^{-itH/\hbar}\chi\rangle$, and $g(t) = \langle\phi| e^{-itH/\hbar}\phi\rangle$.
In order to solve this Volterra type convolution equation numerically, we use a straightforward recursive method based on the discretisation of time.

In our case, anticipating the many-particle picture, it is useful to add an extra conceptual element to the above standard model, namely a ``vacuum'' vector $|0\rangle$ orthogonal to $\mathcal H$, representing the ``absorbed'' state of the system; this ensures that the system (say, a particle) no longer exists in $\mathcal H$ and hence cannot be detected a second time. The absorption action at (random) time $t$ can then be given as $\chi_t\mapsto L\chi_t$ by the operator $L=\sqrt{\gamma}|0\rangle\langle \phi|$, projecting the evolved state $\chi_t$ along the \emph{absorbing} state $\phi\in \mathcal H$ into the vacuum. In this picture, the probability density of the detection time $t$ is given by the norm of the resulting vector; $p(t)=\|L\chi_t\|^2$, and this is clearly consistent with the above description with the same effective Hamiltonian \eqref{Heff}. We note that this supplementing the model with the vacuum vector is a special case of the ``exit space" construction in \cite{Werner1987}.

It is furthermore useful to make a link to open quantum systems: the detection model can be seen as an ``unravelling'' of the master equation
\begin{align*}
\frac{d}{dt}\rho_t &= -\frac{i}{\hbar} [H, \rho_t] +\gamma \langle \phi|\rho_t|\phi\rangle |0\rangle\langle 0| \\
& -\frac{\gamma}{2} (|\phi\rangle\langle\phi|\rho_t+\rho_t|\phi\rangle\langle \phi|),
\end{align*}
which is of the well-known Sudarshan-Gorini-Kossakowski-Lindblad form \cite{Lindblad1976, Gorini1976}, with the single Lindblad operator $L$ introduced above. In this picture, the absorption is a (true) dynamical evolution of an open quantum system with the Hilbert space $\mathcal H\oplus {\rm span}\{|0\rangle\}$ consisting of the system plus the vacuum.

In the following subsections we extend this single-particle model to many-particle detection.

\subsection{Fock space framework}

Since we consider many-particle systems, it is natural to use the Fock space formalism, see for instance \cite{Bratteli1997}, where we restrict to the Bosonic case as discussed in the introduction. We have
\begin{equation}\label{fockspace}
\mathfrak H=\bigoplus_{N=0}^\infty \mathcal H_B^{(N)},
\end{equation}
where $\mathcal H_B^{(N)}$ is the Bosonic (symmetric) subspace of the tensor product $\mathcal H^{\otimes N}$, where $\mathcal H$ is the single-particle Hilbert space. At this stage $\mathcal H$ could be arbitrary (i.e. either finite or infinite-dimensional), but our main results involve a free particle in one dimension, in which case $\mathcal H=L^2(\mathbb R,dx)$, identified as the position space. Each single particle pure state $\phi$ can be identified with a single mode, and the Bosonic creation and annihilation operators $a(\phi)$ corresponding to the different modes satisfy the CCR relation $a(\phi)^\dagger a(\phi') =\langle \phi|\phi'\rangle$. We note \cite{Bratteli1997} that a single application of $a(\phi)$ corresponds to the annihilation of one particle from the system, and accordingly it maps each $N$-particle sector $\mathcal H^{(N)}_B$ into $\mathcal H^{(N-1)}_B$, acting on product states as $a(\phi)|\chi^{\otimes N}\rangle=\sqrt{N}\langle \phi|\chi\rangle |\chi^{\otimes (N-1)}\rangle$; this interpretation will be crucial below in the description of particle absorption. Recall also that in the relevant case $\mathcal H=L^2(\mathbb R,dx)$, we can define the infinitesimal versions $a(x)$ corresponding to the spatial degree of freedom, and we can then expand $a(\phi)=\int \phi(x) a(x)dx$ (in a suitable distributional sense) for any sufficiently smooth normalisable state $\phi\in \mathcal H$.

Finally, we need some basics of 2nd quantisation. For a single-particle operator $A$ (generating a contraction semigroup), we can define the corresponding many-particle version as $q(A)=\oplus_{N=0}^\infty q^{(N)}(A)$, where each $N$-particle contribution is given by
$$q^{(N)}(A)= A\otimes \id \otimes \cdots \otimes\id+ \id \otimes A\otimes \cdots\otimes \id +\cdots +\id\otimes\cdots \otimes A.$$
This, in turn, can be 2nd quantised into the many-particle contraction semigroup $$t\mapsto e^{-it q(A)/\hbar} = \oplus_{N=0}^\infty (e^{-it A/\hbar})^{\otimes N}.$$
We recall, in particular, that the unitary evolution generated by any single-particle Hamiltonian can be 2nd quantised in this way into a unitary evolution of the many-particle system.

\subsection{Many-particle detection}

We now second-quantise the single particle detection model in Subsection \ref{subsec:sp}. First of all, the free evolution of the many-particle system is $t\mapsto e^{-it q(H)/\hbar}$, and we introduce a dissipative part given by the Lindblad operator $L=a(\phi)$ implementing particle absorption by annihilation: it maps each $N$-particle sector into the $N-1$-particle sector, so that (unlike in the single-particle case) the system continues existing but has one particle less. Analogous to the master equation in Subsection \ref{subsec:sp}, we now have the following Lindblad-type model, which has been studied in \cite{Alicki2007, Butz2010}:
\begin{align}\label{master}
\frac {d}{dt} \rho_t &= -\frac{i}{\hbar}[q(H),\rho_t] + \gamma a(\phi) \rho_t a(\phi)^\dagger \nonumber \\
&- \frac \gamma 2(a(\phi)^\dagger a(\phi)\rho_t +\rho_t a(\phi)^\dagger a(\phi) ).
\end{align}
Here $\phi\in \mathcal H$ is the absorption vector introduced in Subsection \ref{subsec:sp}, $\gamma>0$ is again a constant related to the absorption rate, and now $\rho_t$ is the state (density matrix) of the many-particle system at time $t>0$ given some initial state $\rho$. We note here that one could more generally consider absorption in several states \cite{Alicki2007}, but we restrict here to a single one.

Now the unravelling yields the contractive semigroup $\mathcal U_t=e^{-it q(H_{\rm eff})/\hbar}$ generated by 
\begin{equation}\label{effFock}
q(H_{\rm eff})=q(H) - i\frac{\hbar}{2}\gamma a^\dagger(\phi)a(\phi),
\end{equation}
which is easily seen to be the multi-particle version of the effective single-particle Hamiltonian \eqref{Heff}. In fact, by restricting \eqref{effFock} to the single-particle subspace we obtain exactly \eqref{Heff}. Furthermore, we have $\mathcal U_t = \oplus_{N=0}^\infty U_t^{\otimes N}$ where $U_t$ is generated by $H_{\rm eff}$ as before.

Analogous to the single-particle case, the probability of detection before time $t$ is $1- {\rm tr} [\mathcal U_t\rho \mathcal U_t^\dagger]$, and a calculation similar to the one leading to \eqref{singleprob} now gives the relation
$$
\frac{d}{dt} {\rm tr}[\mathcal U_t \rho \mathcal U_t^\dagger]=-\gamma {\rm tr}[J_t \rho J_t^\dagger],
$$
where $J_t = a(\phi) \mathcal U_t$. We therefore obtain
the probability density
$$p_1(t_1)=\gamma {\rm tr}[J_{t_1} \rho J_{t_1}^\dagger]$$ of the first detection time. However, after the first detection, only one particle has been absorbed, so we can continue the process. Here we recall the standard theory of conditional quantum state transformations, or instruments \cite{Davies1976, Barchielli1986}, according to which the density matrix $\tilde \rho_{t_1} =J_{t_1} \rho J_{t_1}^\dagger/p_1(t_1)$ represents the conditional state of the system given that the first detection happened at time $t_1$. Therefore, the corresponding conditional probability density of the next detection at $t_2$ is ${\rm tr}[J_{t_2-t_1} \tilde \rho_{t_1} J_{t_2-t_1}^\dagger]$, so by multiplying it by $p_1(t_1)$ we obtain the joint probability density for $(t_1, t_2)$ as
$$
p_2(t_1,t_2) = \gamma^2{\rm tr}[J_{t_2-t_1} J_{t_1} \rho_0 J_{t_1}^\dagger J_{t_2-t_1}^\dagger].
$$
Similarly, we easily obtain by induction the following form of the joint probability density of $n$ subsequent detections at times ${\bf t}=(t_1,\ldots t_n)$:
\begin{align}\label{detectiontimes}
p_n({\bf t})=\gamma^n{\rm tr}[J_{{\bf t}} \rho J_{\bf t}^\dagger] = \gamma^n{\rm tr}[J_{\bf t}^\dagger J_{\bf t}\rho],
\end{align}
where we have denoted 
$$J_{{\bf t}}= J_{\Delta t_n} J_{\Delta t_{n-1}} \cdots J_{\Delta t_2}J_{\Delta t_1},$$
with $\Delta t_i = t_i-t_{i-1}$ are the time intervals between the arrivals.
Note that the integral of $p_n({\bf t})$ over the simplex $\{ (t_1,\ldots, t_n)\mid t_1<t_2<\cdots < t_n\}$ represents the probability to eventually obtain at least $n$ detections. Here we stress that this probability is not necessarily equal to one.

We remark here that each distribution is obtained by fixing the number, $n$, of detections. The collection of all these distributions (for $n\in \mathbb N$) can be seen (somewhat more abstractly) as specifying the \emph{point process} \cite{Daley1988, Snyder1991} that describes the entire detection statistics. However, rather than building an abstract formalism, we work with concrete special cases with transparent physical interpretation. These will be considered in the next subsection.

\subsection{Arrival time distributions}
The task is now to compute \eqref{detectiontimes} for suitable classes of many-particle source states.     First of all, we restrict entirely to the case where each particle has the same initial wave function $\chi$. 
There are, then, three basic types of many-particle states which we will look at in the following, corresponding to how the varying particle number is treated: fixed particle number, coherent superposition, and a classical mixture. We will show that they will all depend on the effective single-particle evolution $\chi_t = U_t\chi$.

\subsubsection{Fock states}
Consider the \emph{Fock state}
$$
\rho =|\chi^{\otimes N}\rangle\langle \chi^{\otimes N}|,
$$
where $N$ is fixed, so that the state lies in the $N$-particle sector of the Fock space. We now evaluate the density $p_n({\bf t})$, for each fixed number $n$ of arrivals.
Noting first that by the definition of $J_t$, we get
\begin{equation*}
J_t|\psi^{\otimes N}\rangle = a(\phi)|(U_t\psi)^{\otimes N}\rangle=\sqrt{N} \langle \phi| U_t\psi\rangle |(U_t\psi)^{\otimes N-1}\rangle
\end{equation*}
for any $\psi$ and $t$. Consider now two detections at ${\bf t}=(t_2,t_1)$. Using the semigroup property $U_{\Delta t_2}U_{t_1}=U_{t_2}$ and the definition $J_{{\bf t}}=J_{\Delta t_2} J_{t_1}$, we get
\begin{align*}
J_{\bf t}|\chi^{\otimes N}\rangle &= \sqrt{N(N-1)}\langle\phi|\chi_{t_1}\rangle \langle \phi|U_{\Delta t_2}\chi_{t_1}\rangle |(U_{\Delta t_2}\chi_{t_1})^{\otimes N-2}\rangle\\
&= \sqrt{N(N-1)} \langle \phi|\chi_{t_1}\rangle \langle \phi|\chi_{t_2}\rangle |\chi_{t_2}^{\otimes (N-2)}\rangle.
\end{align*}
Continuing by induction we similarly obtain for $n$ detections the form
$$
J_{\bf t} |\chi^{\otimes N}\rangle = \sqrt{\frac{N!}{(N-n)!}} \left(\prod_{i=1}^{n} \langle \phi|\chi_{t_i}\rangle\right)\, |\chi_{t_n}^{\otimes (N-n)}\rangle
$$
when $n\leq N$, and $J_{\bf t} |\chi^{\otimes N}\rangle =0$ otherwise. Here we have used the convention $|\chi_{t_n}^{\otimes 0}\rangle :=|0\rangle$, the Fock vacuum. Hence
\begin{align*}
p_n({\bf t}) = \gamma^n\|J_{\bf t}\chi^{\otimes N}\|^2= \frac{N!\gamma^n}{(N-n)!} \prod_{i=1}^{n} |\langle \phi|\chi_{t_i}\rangle|^2 \|\chi_{t_n}\|^{2(N-n)}
\end{align*}
for $n\leq N$ and zero otherwise. We now define $\omega(t) = N \gamma |\langle \phi|\chi_t\rangle|^2$ and $\Omega(t) = \int_0^t \omega(t') dt'$. 
Noting that $\frac{d}{dt} \|\chi_t\|^2 =-\gamma|\langle \phi|\chi_t\rangle|^2$, we have $\|\chi_t\|^2 = 1-\Omega(t)/ N$. Here it is important to keep in mind that as $\|\chi_t\|^2 \ge 0$ we get $\Omega(t) \le N$ for all $t$, that is, $\Omega(t)$ is bounded by the particle number. With these convenient notations it follows that the arrival time distribution is
\begin{align}\label{fockarrivals}
p_n({\bf t})= \frac{N!}{(N-n)!} \, \left(1-\frac{\Omega(t_n)}{N }\right)^{N-n}\prod_{i=1}^n \frac{\omega(t_i)}{N},
\end{align}
for $n\leq N$ and zero otherwise.
We can compactly rewrite it in the form
\begin{equation}\label{arrivaltimedistributionold}
p_{n}({\bf t}) = (-1)^n F^{(n)}(\Omega(t_n))\prod_{i=1}^n\omega(t_i),
\end{equation}
where $F^{(n)}(\Omega) = (-1)^n\frac{N!}{N^n(N-n)!} \, \left(1-\frac{\Omega}{N}\right)^{N-n}$ is the $n$th derivative of the function $F(\Omega) = (1-\Omega/N)^N$ for $\Omega\leq N$ and $F(\Omega)=0$ for $\Omega>N$. We remark that in the framework of point processes, this type of distribution defines a \emph{multinomial process} with intensity $\omega(t)/N$ \cite{Snyder1991}.

\subsubsection{Coherent states}

We now consider the \emph{coherent} states
\begin{equation}\label{coherentstates}
\rho = |\Psi\rangle\langle \Psi|, \quad \Psi = \bigoplus_{N=0}^\infty \, \frac{\langle N\rangle^{\frac N2} e^{-\langle N\rangle/2}}{\sqrt{N!}} |\chi^{\otimes N}\rangle\in \mathfrak H.
\end{equation}
These are quantum superpositions of Fock states, in such a way that the particle number distribution is Poisson with mean $\langle N\rangle$. Analogous to the Fock state case, we define
\begin{equation}\label{intensity1}
\omega(t) = \langle N\rangle \gamma |\langle \phi|\chi_t\rangle|^2, \quad \Omega(t) = \int_0^t \omega(t') dt',
\end{equation}
noting that this is in fact the same quantity as above, since for the Fock state $\langle N\rangle =N$ where $N$ was fixed in that case.

We again look for the density of the arrival times ${\bf t}=(t_1,\ldots,t_n)$. Proceeding as in the preceding section for each particle number $N$ (and noting that $J_{\bf t}$ maps each $N$ particle sector into the $N-n$ particle sector or to zero if $N<n$), we get
\begin{align*}
p_n({\bf t}) &= \gamma^n\sum_{N=n}^\infty \frac{\langle N\rangle^N e^{-\langle N\rangle}}{N!} \|J_{\bf t}\chi^{\otimes N}\|^2\\
&= \prod_{i=1}^n \left(\frac{\omega(t_i)}{\langle N\rangle}\right)\sum_{N=n}^\infty \frac{\langle N\rangle^{N}e^{-\langle N\rangle}}{(N-n)!} \left(1-\frac{\Omega(t_n)}{\langle N\rangle}\right)^{N-n}\\
&= \frac{\prod_{i=1}^n \omega(t_i)}{\langle N\rangle^{n}}\sum_{N=0}^\infty \frac{\langle N\rangle^{(N+n)}e^{-\langle N\rangle}}{N!} \left(1-\frac{\Omega(t_n)}{\langle N\rangle}\right)^{N}\\
&= e^{-\Omega(t_n)}\prod_{i=1}^n \omega(t_i),
\end{align*}
which has the form of a \emph{Poisson process} with intensity $\omega(t)$ \cite{Snyder1991}. We stress that it is again of the form \eqref{arrivaltimedistributionold}, but now with the function $F(\Omega)= e^{-\Omega}$.

\subsubsection{Quasi-free states}

Finally, we study specific classical mixtures of Fock states, generated via a simple Bernoulli trial (coin toss) sequence: at each trial a particle is created with probability $\beta$, and the trials are repeated until the first failure. (One can think of an ``oven" emitting particles, in which case $\beta$ would be related to temperature.) Hence the particle number distribution is Geometric, $\mathbb P(N=k) = (1-\beta)\beta^k$, with the mean particle number $\langle N\rangle = \beta/(1-\beta)$, and the resulting state is
\begin{equation}\label{quasifreestate}
\rho = \frac{1}{1+\langle N\rangle}\bigoplus_{N=0}^\infty \left(\frac{\langle N\rangle}{1+\langle N\rangle}\right)^N |\chi^{\otimes N}\rangle\langle \chi^{\otimes N}|.
\end{equation}
This is a particular case of a \emph{quasi-free} state, also sometimes called thermal or chaotic state; see e.g. \cite{Bratteli1997, Kiukas2013} for a general description of such states.

Defining again the intensity $\omega(t)$ by \eqref{intensity1} and proceeding as in the coherent case, we get
\begin{align*}
p_n({\bf t}) &= \frac{1}{1+\langle N\rangle}\sum_{N=n}^\infty \left(\frac{\langle N\rangle}{1+\langle N\rangle}\right)^N \|J_{\bf t}\chi^{\otimes N}\|^2\\
&= \frac{n! \prod_{i=1}^n \omega(t_i)}{n (1+\langle N\rangle)^{n+1} q^n} \sum_{N=n}^\infty N\binom{N-1}{n-1}q^n\left(1-q\right)^{N-n},\\
&= \frac{n!\prod_{i=1}^n \omega(t_i)}{n q^n (1+\langle N\rangle)^{n+1}}  \frac nq= \frac{n!}{(1+\Omega(t_n))^{n+1}}\prod_{i=1}^n \omega(t_i)
\end{align*}
where $q=(1+\Omega(t_n))/(1+\langle N\rangle)$, and we have recognised the sum on the second line as the expected value of a Negative Binomial distribution. Again we have the form \eqref{arrivaltimedistributionold}, now with $F(\Omega) = (1+\Omega)^{-1}$. We remark that point processes of this type are called \emph{Negative Binomial} \cite{Gregoire1983}, or more generally \emph{mixed Poisson} \cite{Snyder1991}.

\subsection{Summary of arrival time distributions}

So far, we have derived a formula for the arrival time probability distribution $p_{n}({\bf t})$ of observing $n$ subsequent detections at times $0<t_1<t_2<\cdots<t_n$. We have studied this for three basic types of many-particle states, corresponding to how the varying particle number is treated: fixed particle number, coherent superposition, and a classical mixture. In all the three cases, the distribution turned out to be of the form
\begin{equation}\label{arrivaltimedistribution}
p_{n}({\bf t}) = F_n(\Omega(t_n))\prod_{i=1}^n\omega(t_i),
\end{equation}
with $F_n=(-1)^n F^{(n)}$ and where $F^{(n)}$ is the $n$th derivative of the function
\begin{eqnarray}
F(\Omega) = \left\{
\begin{array}{ll}
\left(1-\frac{\Omega}{N}\right)^{N} \text{ for } \Omega<N & \mbox{ (Fock state)}\\
e^{-\Omega} & \mbox{ (Coherent state)}\\
(1+\Omega)^{-1}& \mbox{ (Quasi-free state)}
\end{array}\right. .
\end{eqnarray}
Here, the ``intensity'' function and its integrated version are given by
\begin{align}\label{intensity}
\omega(t) &= \langle N \rangle \gamma|\langle \phi|\chi_{t}\rangle|^2, & \Omega(t)&=\int_{0}^t\omega(t')dt',
\end{align}
where $\langle N\rangle$ is the mean particle number, $\chi_t = U_t\chi$ is the effective (contractive) evolution of the single-particle source state $\chi$.
It is important to note that we have
$\|\chi_t\|^2 = 1-\Omega(t)/ \langle N \rangle$ and therefore (as $\|\chi_t\|\leq 1$) we obtain $\Omega(t) \le \langle N\rangle < \infty$ for all $t$.
Note also that $\omega(t)$ has the unit of 1$/$time, so that the integrated intensity $\Omega(t)$ is dimensionless.

We observe that in each case the detection time density depends explicitly on the intensity function $\omega(t)$, which in turn is essentially (i.e up to scaling by the mean particle number) equal to the single-particle arrival probability \eqref{singleprob}.
However, we stress that the probability distribution for the first arrival in a many-particle system depends explicitly on the mean particle number of the system; in particular, it is not equal to the arrival time distribution of a single particle. Therefore, the above many-particle model is necessary to describe the arrival time detections.

\subsection{Consideration of the no-event probability}\label{subsec:noevent}

We will now examine in detail the normalisation of the detection time distributions $p_n$. We stress that the normalisation is not necessarily one. The physical meaning of this is that, one the one hand, with each particle there is a nonzero probability of never getting detected as the detector is not perfect, and one the other hand, the source state might not have sufficient number of particles. As it will turn out, this crucially affects the single-particle information we can extract from the detection statistics.

Given that the full range of the arrival time sample ${\bf t}$ is the simplex $\Delta_n =\{ {\bf t}\mid 0<t_1<t_2<\cdots<t_n\}$, we find the total probability to obtain at least $n$ detections:
\begin{equation}\label{totalprob}
p_n^{\rm tot} := \int_{\Delta_n} p_n({\bf t}) d{\bf t}=\frac{1}{(n-1)!}\int_0^{\Omega(\infty)} F_n(\Omega) \,\Omega^{n-1}d\Omega.
\end{equation}
Using properties of the functions $F_n$ we obtain (see Appendix \ref{app:noevent})
\begin{align}\label{totaldetprob}
p_n^{\rm tot} &= \begin{cases}
1- \sum\limits_{k=0}^{n-1} \frac{F_n(\Omega(\infty))}{k!} \Omega(\infty)^k, & \text{ if } \Omega(\infty) <\infty\\
1, & \text{ if } \Omega(\infty) =\infty.
\end{cases}
\end{align}
This shows that $p_n^{\rm tot}<1$ when $\Omega(\infty)$ is finite, and all particles are detected (only) when $\Omega(\infty) =\infty$. Importantly, the latter case never happens in the above models because $\Omega(t)\leq \langle N\rangle <\infty$ as noted above. However, the case $\Omega(\infty)=\infty$ is nevertheless relevant, as it will be obtained in the next Section, at the interesting limit of the plane wave beam where $\langle N\rangle \rightarrow \infty$, see Fig. \ref{fig:context}(a) and the discussion in the introduction.

One important consequence of \eqref{totaldetprob} is that in all the above Fock space models we always have
\begin{equation}\label{pntotlimit}
\lim_{n\rightarrow\infty} p_n^{\rm tot} = 0.
\end{equation}
This is trivial for the Fock state case, and follows for the coherent and quasi-free cases because $F_n=(-1)^n F^{(n)}$, so the sum in \eqref{totaldetprob} is the partial sum of the Taylor series of the function $F$ around the point $\Omega(\infty)$ evaluated at zero, and hence converges due to analyticity to $F(0)=1$ as $n\rightarrow \infty$.

This observation suggests that the beam case will have very different statistical properties, and we will indeed later see how this is reflected in the Fisher information.

\subsection{Application: massive free particles in 1D}\label{subsec:application}
We now turn to the case of massive Bosons moving freely and detected by the above detector model. We restrict to the one-dimensional case, which already shows the relevant features. The single-particle Hilbert space is $\mathcal H=L^2(\mathbb R, dx)$ in the position representation, and for any $\chi\in \mathcal H$ we let $\hat \chi$ denote the corresponding momentum space wavefunction given by the Fourier transform $\hat\chi(p) = (2\pi\hbar)^{-\frac 12}\int e^{-ipx/\hbar}\chi(x)\, dx$. The Hamiltonian is $$H=-\frac{\hbar^2}{2m} \frac{d^2}{dx^2}.$$ 
Let $\gamma_\epsilon = \frac{a}{2\epsilon \sqrt{2\pi}}$ and for the absorbing state $\phi$ we take a Gaussian centered at the detector position which we choose $x=0$:
\begin{align}\label{gaussiandetector}
     \phi_\epsilon(x) &= \frac{e^{-\frac 14 x^2/\epsilon^2}}{\sqrt{\epsilon}(2\pi)^{\frac 14}},
\end{align}
where $\phi_\epsilon$ is normalised (in the $L^2$-norm) with width $\epsilon>0$ in position space, and $a>0$ is a constant.
For the single-particle source state $\chi$ we take a Gaussian on the left-hand side of the detector moving towards it, i.e. in the momentum space, we have
\begin{align}\label{initialgaussian}
\hat\chi(p) = \frac{e^{-\frac{1}{4\Delta p^2}(p-p_0)^2-ipx_0/\hbar}}{\Delta p^{\frac 12}(2\pi)^{\frac 14}},
\end{align}
where $\Delta p>0$, $p_0>0$ and $x_0<0$ are width, initial momentum and position, respectively. 

\begin{figure}[th!]
\begin{center}
\includegraphics[scale=0.7]{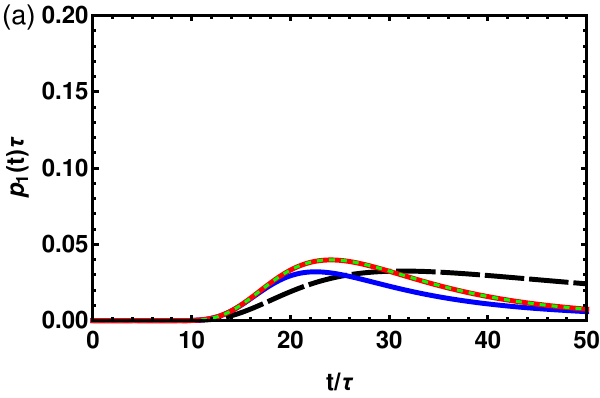}\\
\includegraphics[scale=0.7]{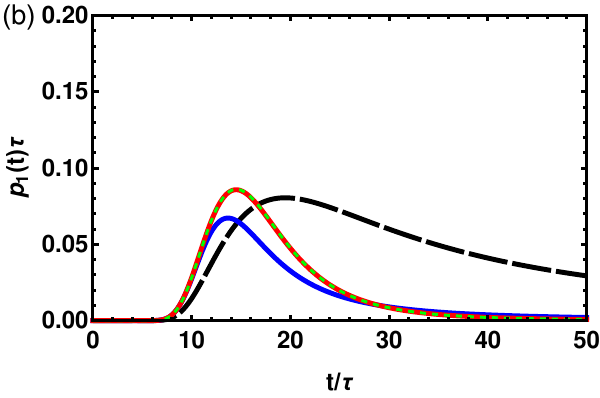}\\
\includegraphics[scale=0.7]{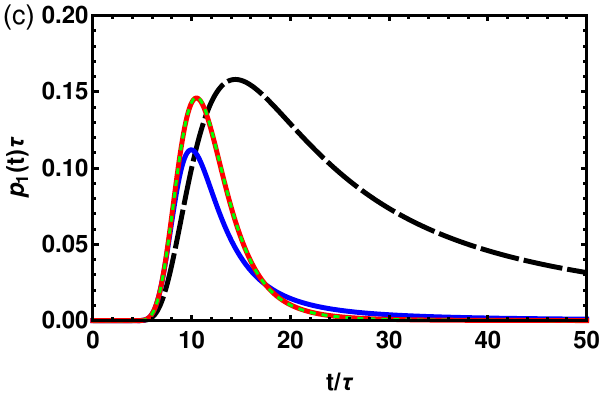}
\end{center}
\caption{\label{fig:convergence} Probability distributions $p_1(t)$ for the first arrival times:
single particle case (black, dashed line, reduced by a factor of $2$); many-particle case with average particle number $\langle N\rangle=100$:
Fock (green, dotted lines, on top of red lines), coherent (red, solid lines), and quasi-free (blue, solid lines) beams. Finite beam with initial one-particle state \eqref{initialgaussian} with $p_0/\bar p=1$, $x_0/l=-20$, $\Delta p^2/\bar p^2=0.5$, $m l/(\bar p \tau) = 1$.
Detector state \eqref{gaussiandetector} with $a \tau/l =0.1$, three detector state widths: (a) $\epsilon/l = 1$, (b) $\epsilon/l =0.5$, and (c) the delta detector limit $\epsilon = 0$.} 
\vspace{-0.5cm}
\end{figure}

In order to illustrate the resulting arrival time distributions, we consider in Fig. \ref{fig:convergence} the probability distribution $p_1(t)$ for the first arrival time for the three types of many-particle states; this is given as a special case
of eq. \eqref{arrivaltimedistribution} by setting $n=1$, so that the vector $\bf t$ just becomes a single variable $t$. 
In the figure we use $x_0/l =-20$, $p_0/\bar p=1$, $a\tau/l=0.1$, where $\tau$ is the time unit, $l$ the length unit and $\bar p= \hbar/l$ the momentum unit (chosen such that the value of $\hbar$ is one in these units). Then the mass unit is $\bar p \tau/l$.

The green, dotted line corresponds to the Fock state,
the red line to the coherent case, and the blue line to the quasi-free state. In all cases, we are considering $\langle N\rangle=100$. For this particle number, the coherent and Fock results are essentially indistinguishable; it is well known that the Fock state and the coherent state converge to each other for large particle number and this can be seen also here. Different widths of the absorbing state are considered in Fig. \ref{fig:convergence} parts (a) ($\epsilon/l=1$) and (b) ($\epsilon/l=0.5$).

For comparison, we have also included the single-particle case (see Subsection \ref{subsec:sp}) in the figures: the black, dashed line is the single-particle arrival time probability density eq. \eqref{singleprob} (reduced by a factor of $2$). Finally, to get an intuitive picture of the time scale, note that the expected arrival time according to classical physics would be centered around $t_C/\tau = 20$ with the above choices of parameters.

We observe that with weak detectors (large $\epsilon$), the peak of the arrival time distribution occurs after the classical time, while for stronger detectors (including the limiting delta detector), it occurs before the classical value.
We see a clear effect of the different many-particle states in comparison to the single-particle result; the maximum appears earlier in the many-particle cases.

\section{Limit models: Delta detector and particle beams}\label{sec:deltabeamlimits}

This section is devoted to a study of two idealised limiting cases in the setting of the application in subsection \ref{subsec:application}. First of all, we consider detection at a single point $x=0$ in the position space. This is in some sense the most canonical choice for an ideal detector placed at the origin without any further details on the absorbing state. Note that such a ``Dirac delta" type detector is not included in the above description based on a normalised absorbing state $\phi$. However, we will show below that it can be derived using absorbing states approximating the Dirac delta distribution at $x=0$ with the normalisation depending on the mean particle number. Secondly, we consider a similar limit involving source states $\chi$ localised in the \emph{momentum} space around a single value $p_0$, such that the resulting system, containing infinite number of particles, corresponds to a spatially uniform ``plane wave beam". 

\subsection{Dirac delta detector model}

We treat point detection as a limiting case of our model, as follows. Let $a>0$ be a constant describing the strength of the detection; if we choose absorbing states $\phi_\epsilon$ for each $\epsilon >0$, such that
\begin{align}\label{delta}
\sqrt{\gamma_\epsilon}\phi_\epsilon(x) &\rightarrow \sqrt{a}\delta(x), \quad \text{ as }\epsilon\rightarrow 0
\end{align}
(that is, $\sqrt{\gamma_\epsilon}\langle \phi_\epsilon|\psi\rangle \rightarrow \sqrt{a} \langle 0|\psi\rangle$ for any $\psi$ continuous at $0$), then, heuristically, we indeed obtain an imaginary delta-potential: $H_{\rm eff}\rightarrow H-i\frac {\hbar}{2} a \delta(x)$. In the process of passing to the limit, $\gamma_\epsilon$ in the original master equation must tend to infinity (as $\phi_\epsilon$ is normalised), while the annihilation operator $a(\phi)=\int \phi_\epsilon(x) a(x) dx$ shrinks towards $a(0)$, corresponding to annihilation at a single spatial mode $x=0$. Here we emphasise that we do not need to discuss a mathematically rigorous formulation of these operator limits, as it suffices to show that the intensity function \eqref{intensity}, and hence the relevant arrival time distributions \eqref{arrivaltimedistribution}, converge to a well-defined limit which does not depend on the choice of the sequence of the absorbing states $\phi_\epsilon$. Note that in the numerical example below, we implement \eqref{delta} using the Gaussian \eqref{gaussiandetector} where the parameterisation by $a$ and $\epsilon$ is specifically chosen so that \eqref{delta} holds.

We now show the required convergence of the intensity function under the assumption that the Fourier transform $\hat\chi$ of the initial source state $\chi$ is integrable. We consider the integral equation \eqref{integraleq}, where the initial function $h_{0,\epsilon}(t)$ and the kernel $g_\epsilon(t)$ now depend on $\epsilon$. First note that by using \eqref{delta}, together with the kernel of the free propagator $\langle x|e^{-itH/\hbar} x' \rangle = \sqrt{\frac{m}{2\pi i \hbar t}} \exp\left(-\frac{m (x-x')^2}{2i\hbar t}\right)$, we get
\begin{align*}
\sqrt{\gamma_\epsilon}h_{0,\epsilon}(t)&= \sqrt{\gamma_\epsilon}\langle \phi_\epsilon|e^{-itH/\hbar}\chi\rangle \rightarrow \sqrt{a} \langle 0|e^{-itH/\hbar}\chi\rangle\\
\gamma_\epsilon g_\epsilon(t)&=\gamma_\epsilon\langle \phi_\epsilon |e^{-itH/\hbar}\phi_\epsilon\rangle\rightarrow \frac{a\sqrt{m}}{\sqrt{2\pi i\hbar t}},\quad \text{ as }\epsilon\rightarrow 0
\end{align*}
where in $\sqrt{i}=e^{i\pi/4}=\frac{1}{\sqrt{2}}(1+i)$ we use the principal branch of the square root, and the assumption on $\hat\chi$ ensures that $e^{-itH/\hbar}\chi$ is continuous so that \eqref{delta} applies. Now, if we multiply both sides of the equation \eqref{integraleq} by $\sqrt{\gamma_\epsilon/a}$ we obtain the determining integral equation for the function $f_\epsilon(t) = \sqrt{\gamma_\epsilon/a}\langle \phi_\epsilon|\chi_{\epsilon,t}\rangle$ (where $\chi_{\epsilon,t}$ implicitly depends on $\epsilon$ through that equation). But using the above limits we observe that both the initial function and the kernel of that integral equation converge, leading to the equation 
\begin{equation}\label{renewal}
f(t) = f_{\rm free}(t) - \frac{d}{\sqrt \pi}\int_0^t \frac{f(s)}{\sqrt{t-s}} ds,
\end{equation}
where $f(t)$ is the solution, $f_{\rm free}(t):=(e^{-it H/\hbar }\chi)(0)$ is known (the freely evolving wavefunction at the origin), and $d=\frac{a\sqrt m}{2\sqrt{2i\hbar}}=\frac{a \sqrt m}{4\sqrt \hbar}(1-i)$ is a constant with $1/d^2$ having the unit of time. Therefore, it follows that the solution $f_\epsilon(t)$ converges to $f(t)$ (see Appendix \ref{app:deltaequation} for a rigorous proof and a consideration of the type of convergence), and hence we can take the limit $\epsilon\rightarrow 0$ in the intensity function \eqref{intensity} to get 
\begin{equation}\label{omegadelta}
\omega(t)=a\langle N\rangle|f_\epsilon(t)|^2\rightarrow \omega_{{\rm delta}}(t):=a \langle N\rangle|f(t)|^2.
\end{equation}
Hence, the detection process is well-defined at the delta detector limit, and is completely determined by the integral equation \eqref{renewal}. The equation can be solved by standard Laplace methods (see Appendix \ref{app:deltaequation}); the general solution is given by 
\begin{equation}\label{gensol}
f(t) = \int_{-\infty}^\infty f_p(t)\, \hat\chi(p) dp,
\end{equation}
where
\begin{equation}\label{deltapsol}
f_p(t) = \frac{1}{\sqrt{2\pi\hbar}}\left(T_p + \widetilde R_p(t)\right) e^{-it p^2/(2m\hbar)},
\end{equation}
where we recognise
\begin{eqnarray}
T_p = |p|/(|p|+\tilde \alpha)
\label{deltatransmission}
\end{eqnarray}
as the transmission amplitude of the imaginary potential $-i\frac{\hbar}{2} a \delta(x)$, and the remainder term is
\begin{align}
\widetilde R_p(t) = \frac{\tilde\alpha}{p^2-\tilde\alpha^2}&\Big(|p|\,{\rm erfc}(\tfrac{|p|}{\sqrt{2m\hbar}}e^{-i\frac{\pi}{4}}\sqrt{t})\nonumber\\
&- \tilde\alpha \,e^{it\frac{p^2-\tilde\alpha^2}{2m\hbar}} {\rm erfc}(\tfrac{\tilde\alpha}{\sqrt{2m\hbar}} e^{-i\frac{\pi}{4}}\sqrt{t})\Big).
\label{deltareflection}
\end{align}
Here $\tilde \alpha = a m /2$ has the units of momentum.
Hence $f_p(t)$ splits into a stationary part and the transient part $\widetilde R_p(t)$. We can now use this result to find the intensity function \eqref{omegadelta} for a given source state $\chi$. This function, then, determines the arrival time distributions \eqref{arrivaltimedistribution} for the delta detector.

\subsection{Spatially uniform particle beams}\label{sec:beam}

Our next aim is to study arrival time distributions for particles sources with definite momentum and uniform spatial particle density. Such a source state has the interesting feature that position measurements do not yield any information on the single-particle momentum parameter $p_0$. In contrast, the latter can be estimated from arrival time data, as we will see below in Section \ref{sec:metrology}.

To be precise, we call a source state a \emph{(plane wave) beam}, if it has the following two properties:
\begin{enumerate}
\item[(i)] Each particle has definite momentum $p_0$. This is the relevant single-particle parameter.
\item[(ii)] The spatial particle density (mean number of particles per unit length) is a constant $r_0$.
\end{enumerate}
We stress that the idea is to \emph{define} the beam source by the above two properties, that is, only in terms of the \emph{position} distribution of the particles on bounded intervals, hence independently of the measurements we are interested in. We then subsequently \emph{prove} that the arrival time distributions are well-defined for the beam.

We remark that this notion differs from \emph{temporal} stationarity defined through commutativity of a (mixed) single-particle state with the system Hamiltonian, as in e.g. \cite{Kiukas2013}. The present definition is static, i.e. describes the state of the system in one time instant, independently of any dynamics.

Clearly, such a beam state cannot be described within the Fock space, as the number of particles is necessarily infinite. There are various ways to construct an abstract description of infinite systems \cite{Bratteli1997, Kiukas2013}; here we instead follow a very concrete approach, involving a limit of states in the Fock space which does not converge within it, but nevertheless leads to convergent position distributions for any finite interval. 

Accordingly, consider any finite interval $I$ in the single-particle position space, and let $Q_I$ be the corresponding projector $\psi\mapsto \chi_{I}\psi$ where $\chi_I(x)=1$ for $x\in I$ and zero otherwise. Then the 2nd quantised operator $q(Q_I)$ is the corresponding counting observable, that is, the mean number of particles in the interval $I$ in a given many-particle state $\rho$ is $\langle q(Q_I)\rangle_{\rho}={\rm tr}[q(Q_I)\rho]$. We can now show (see Appendix \ref{app:spatial}) that in each of the cases considered above (Fock, coherent, quasi-free), we have $\langle q(Q_I)\rangle= \langle N\rangle \int_I |\langle\chi|x\rangle|^2dx$, where $\chi$ is the single-particle source state. Therefore, the spatial particle density is
\begin{equation}\label{particledensity}
r(x) = \langle N\rangle |\langle x|\chi\rangle|^2.
\end{equation}
Let now $\chi_\nnx$ be a single-particle state parameterised by the mean particle number $\nn$, where $\nn\rightarrow \infty$ should describe the plane wave limit.
First, to satisfy (i) we need to scale $\chi_\nnx$ so that it approximates the eigenstate of momentum with eigenvalue $p_0$; more specifically, there need to exist a constant $\lambda_\nnx$ (which will tend to infinity for $\nn \to \infty$) in such a way that
\begin{equation}\label{planewave1}
\sqrt{\lambda_\nnx}\,\hat\chi_\nnx(p) \rightarrow \delta(p-p_0)\, \mbox{ for } \nn \to \infty.
\end{equation}
Since $\lim_{\nn \rightarrow \infty} \sqrt{\lambda_\nnx}\int \hat\chi_\nnx(p)dp =1$ is therefore dimensionless, it follows that $1/\lambda_\nnx$ has the units of momentum. 
Next, to satisfy (ii) we want $r_0 = \lim_{\nn\rightarrow \infty} r_\nnx(x)$ to exist and be independent of $x$. 
Given that $\sqrt{\lambda_\nnx}\,\langle x|\chi_\nnx\rangle = \frac{1}{\sqrt{2\pi\hbar}}\int e^{ixp/\hbar} \sqrt{\lambda_\nnx}\,\hat\chi_\nnx(p) dp \rightarrow \frac{1}{\sqrt{2\pi\hbar}} e^{ixp_0/\hbar}$ for each $x$, this follows from the existence of the limit
\begin{eqnarray}\label{planewave2}
r_0 &=& \lim_{\nn\rightarrow \infty} \left(\frac{\nn}{\lambda_\nnx}\right) \left|\langle x|\sqrt{\lambda_\nnx}\chi_\nnx\rangle \right|^2\nonumber\\
&=& 
\lim_{\nn\rightarrow \infty} \nn/(2\pi \hbar\lambda_\nnx).
\end{eqnarray}
Hence the two equations \eqref{planewave1} and \eqref{planewave2} constitute a reasonable definition of a plane wave beam. In Appendix \ref{app:spatial} we show in detail how (the characteristic function of) the full position distribution of the particles in any finite interval converges in the beam limit to a distribution which only depends on $r_0$ (and the interval length), that is, does not contain any single particle information. The limit describes how the particles are spatially distributed in the beam; the distribution is Poisson for both the Fock state and coherent cases, and Geometric for the quasi-free case.

In the numerical example below, we take $\chi_\nnx$ given by \eqref{initialgaussian} with
\begin{eqnarray}
\Delta p_\nnx = \sqrt{\frac{ \pi}{2}} \frac{\hbar r_0}{\nn}.
\label{defpnn}
\end{eqnarray}
Note that in this case $r_\nnx(x_0) = r_0$ has a meaning for all values of $\nn$: in fact, $r_0$ is the maximal spatial particle density of the finite-particle state determined by \eqref{initialgaussian}, for each $\nn$. In the following we therefore simply fix the normalisation constant as
\begin{eqnarray}
\lambda_\nnx = \langle N\rangle/(2\pi \hbar r_0),
\end{eqnarray}
for each $\langle N\rangle$.

\subsection{Arrival time distributions for the beam}

We now show that the beam source specified by the single-particle momentum $p_0$ and the spatial particle density $r_0$, yields well-defined arrival time distributions for the delta detector described above. To that end, let $f^{(\nnx)}(t)$ be the solution \eqref{gensol} for the initial state $\chi=\chi_\nnx$, for each $\langle N\rangle$. From \eqref{omegadelta} we now get
\begin{align*}
\omega_{{\rm delta},\nnx}(t)&:= a  \langle N\rangle |f^{(\nnx)}(t)|^2\\
&= a \langle N\rangle | \int_{-\infty}^\infty f_p(t)\, \hat\chi_\nnx(p) dp |^2\\
&= a \frac{\langle N\rangle}{\lambda_\nnx} \left|\int_{-\infty}^\infty f_p(t)\, \sqrt{\lambda_\nnx} \hat\chi_\nnx(p) dp \right|^2.
\end{align*}
But now it follows from our definition of the beam (\eqref{planewave1} and \eqref{planewave2}) that 
\begin{align*}
\omega_{\rm delta}^{\rm beam}(t)&:=\lim_{\langle N\rangle\rightarrow \infty}\omega_{{\rm delta},\nnx}(t) = 2\pi a r_0 \hbar |f_{p_0}(t)|^2 \nonumber\\
&= a r_0 |T_{p_0}+\widetilde R_{p_0}(t)|^2,
\end{align*}
that is, the intensity function tends to a finite limit, showing that the arrival time distributions are well-defined for the beam. 

It is interesting to consider the small and large time asymptotics of the intensity function. For small $t$, we get
\begin{eqnarray}
\omega_{\rm delta}^{\rm beam}(t) &=& a r_0 \left(1 - \frac{a m}{\sqrt{\hbar m \pi}} \sqrt{t} + \OO(t) \right),\label{omegaapproxsmall}\\
\frac{d\omega_{\rm delta}^{\rm beam}}{dp}(t) &=& a r_0 \left(- \frac{a m p_0}{3 \sqrt{\pi} (\hbar m)^{3/2}} t^{3/2} + \OO(t^2)\right)\label{domegaapproxsmall}\!\!.
\end{eqnarray}
Now we consider the opposite case of large $t$.
The remainder function $\widetilde R_p$ has a power-law decay; indeed, using the asymptotic expansion of the error function \cite{Abramovitz1965} we obtain $\widetilde{R}_p(t)\sim \OO(t^{-\frac 32})$.
We then also get
\begin{eqnarray}
\omega_{\rm delta}^{\rm beam}(t) &=& a r_0 \left(\frac{p_0^2}{(\frac{a m}{2} + p_0)^2} + \OO(t^{-3/2})\right),
\label{omegaapprox}\\
\frac{d\omega_{\rm delta}^{\rm beam}}{dp}(t) &=& a r_0 \left(\frac{a m p_0}{(\frac{a m}{2} + p_0)^3} + \OO(t^{-1/2})\right)\label{domegaapprox}.
\end{eqnarray}
These expansions will be needed in the subsequent section.

\subsection{Limit models: summary and discussion}
In this section, we have shown so far that the intensity function \eqref{intensity}, needed to determine the many-particle arrival time probability density \eqref{arrivaltimedistribution}, can be calculated in the case of a delta detector as
\begin{equation}
\omega_{{\rm delta}}(t)= \frac{a\langle N\rangle}{2\pi\hbar}\left|\int_{-\infty}^\infty \big[T_p + \widetilde R_{p}(t)\big] e^{-it\frac{p^2}{2m\hbar}}\, \hat\chi(p) dp\right|^2,
\label{delta_intensity}
\end{equation}
where $\hat \chi(p)$ is the usual momentum representation of the single-particle source state $\chi$, while $T(p)$ and $\widetilde {R}_p(t)$ are given in \eqref{deltatransmission} and \eqref{deltareflection}, respectively. Note that here we still have $\Omega(\infty) < \infty$ and therefore $p_{n}^{\rm tot} < 1$.

Our second limit involves the source state tending to a spatially uniform beam containing infinite number of particles, defined by \eqref{planewave1} and \eqref{planewave2}. Note that we took this limit after the delta-detector limit. The resulting intensity function is given by
\begin{eqnarray}\label{plane_wave_intensity}
\omega_{\rm delta}^{\rm beam}(t) = a r_0 |T_{p_0} + \widetilde R_{p_0}(t)|^2,
\end{eqnarray}
and the corresponding many-particle arrival time distribution is obtained from \eqref{arrivaltimedistribution} with $\omega=\omega_{\rm delta}^{\rm beam}$. This shows that the delta-detection model is nontrivial even for a spatially uniform beam. Note that the intensity function depends explicitly on $a$ (the detection strength), $p_0$ (the single-particle momentum), and $r_0$ (the spatial particle density), as well as the single-particle mass $m$. These are the only parameters remaining in this limit.

We remark that the arrival time process is clearly not \emph{time-stationary} even after the two idealising limits, as $\omega_{\rm delta}^{\rm beam}(t)$ is not constant for small times. This is interesting given the spatial uniformity of the plane wave beam which is part of our definition. The lack of time-stationarity is of course due to the fact that the absorbing detector is turned on at the initial time $t=0$, making the setting temporally asymmetric. However, for large $t$ we do get from \eqref{omegaapprox} that
\begin{equation}\label{constantomega}
\lim_{t\to\infty} \omega_{\rm delta}^{\rm beam}(t)
=\omega_{\rm delta}^{\rm beam}(\infty) =  
\frac{a r_0p_0^2}{(\frac{a m}{2} + p_0)^2},
\end{equation}
that is, the intensity tends to a constant. This also means that now $\Omega(\infty) = \infty$, which implies that $p_{n}^{\rm tot} = 1$ for all $n$, i.e. all the arrival time distributions are normalised in the beam limit.

\subsection{Limit models: example and discussion}

We begin by considering a numerical example of the distribution of the first arrival density $p_1(t)$ illustrating the passage to delta detector limit, see again Fig. \ref{fig:convergence}. Here the particle number is fixed (so we are not yet taking the beam limit), and we are using the Gaussian absorbing states \eqref{gaussiandetector} with different widths adjusted by $\epsilon$; the delta detector limit $\epsilon=0$ is shown in Fig. \ref{fig:convergence}(c); here we see again that the presence of multiple particles crucially affects even the probability of the first arrival.

\begin{figure}[t]
\begin{center}
\includegraphics[scale=0.65]{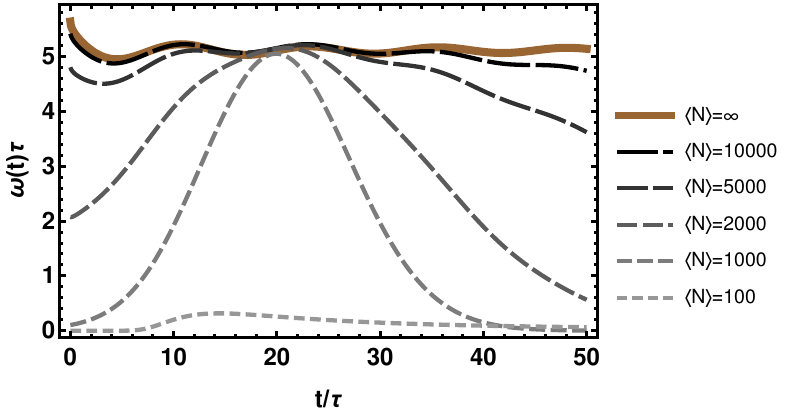}
\end{center}
\caption{\label{fig:planewaveA} Convergence to the beam limit for the delta-detector: intensity function $\omega_{\rm delta}(t)$
for different values of the average particle number $\langle N\rangle$; analytically computed limit $\omega_{\rm delta}^{\rm beam}(t)$ (see \eqref{plane_wave_intensity}; brown, solid line). Spatial particle density $r_0 l=56.42$, detector strength
$a \tau/l=0.1$, the initial states $\chi_\nnx$ are given by \eqref{initialgaussian} using \eqref{defpnn} and $p_0/\bar p=1$, $x_0/l=-20$, $m l/(\bar p \tau) = 1$.
} 
\vspace{-0.5cm}
\end{figure}

As the next step, we are now looking at the same example in the  subsequent beam limit. The convergence of the intensity $\omega_{\rm delta}(t)$ to $\omega_{\rm delta}^{\rm beam}(t)$ is shown in Fig. \ref{fig:planewaveA}.
In the beam limit the large time constant value \eqref{constantomega} (which is $\omega_{\rm delta}^{\rm beam}(\infty) = 5.12$ in this example) is clearly visible with oscillations around it.

\begin{figure}[t]
\begin{center}
\includegraphics[scale=0.65]{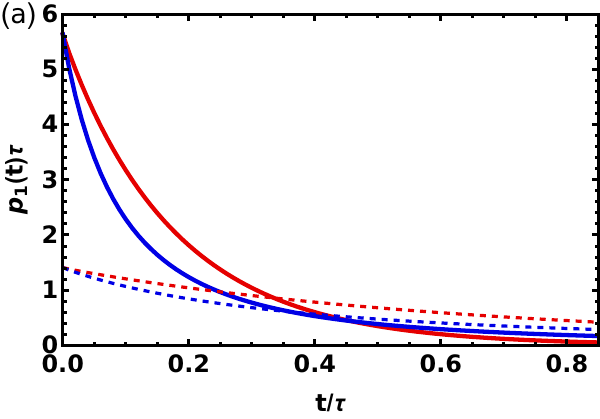}\\
\includegraphics[scale=0.6]{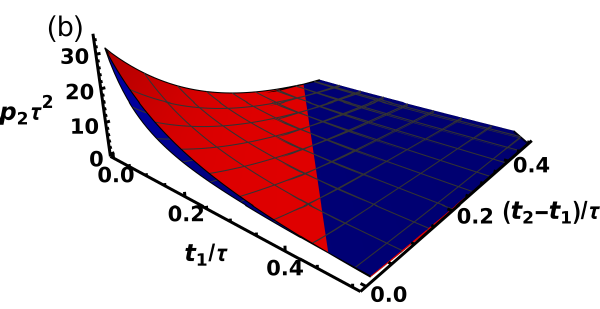}
\includegraphics[scale=0.6]{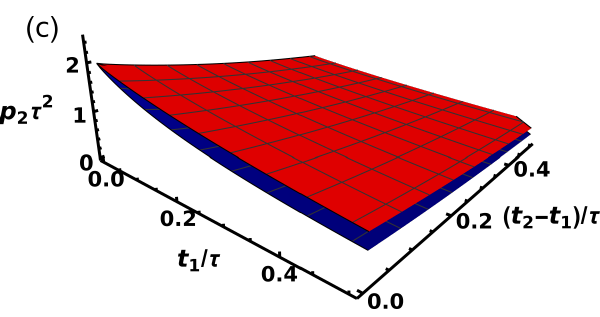}
\end{center}
\caption{\label{fig:planewaveB} 
(a) First-detection probability density $p_1 (t)$,
(b,c) two-detection probability density $p_2 (t_1, t_2)$.
In the delta detector and beam limit case: coherent state (red lines/surfaces), quasi-free state (blue lines/surfaces). Spatial particle densities $r_0 l=56.42$ (solid lines and (b)), $r_0 l=56.42/4$ (dotted lines and (c)). The other values are 
$a \tau/l=0.1$, $p_0/\bar p=1$, $m l/(\bar p \tau) = 1$.
} 
\vspace{-0.5cm}
\end{figure}

Figure \ref{fig:planewaveB} (a) shows the probability density of the first arrival time for the beam with different values of the spatial particle density $r_0$. We observe that the probability density is now peaked at zero instead of a positive value, reflecting the fact that the beam particles are uniformly distributed in space. We also see that the arrival probability for small times is larger for a coherent state (red lines) than for the quasi-free state (blue lines).

It is interesting to consider the analytical form of the first arrival probability density in the beam limit for small times. By expanding $p_1(t)$ into a power series using the form of the $\omega_{\rm delta}^{\rm beam}(t)$ obtained above we get
\begin{eqnarray}\label{firstarrivalexpansion}
p_1(t) &=& r_0 a - \frac{a^2 \sqrt{m} \, r_0}{\sqrt{\pi\hbar}} \sqrt{t} \nonumber\\
&&+  \frac{a^2 r_0^2 (a m/(\hbar r_0) - 3\pi \pm \pi)}{2\pi} t + \OO (t^{3/2})
\label{series}
\end{eqnarray}
with $+$ for the coherent state and $-$ for the quasi-free state.
From this, it follows that $p_1(0)=r_0 a$, i.e. $p_1(0)$ increases with increasing particle density $r_0$. This is expected since for a large particle density the probability density for a first detection immediately at $0$ should be larger. As the distribution is normalised to $1$, the probability distribution also gets more and more centered around $t=0$ for increasing $r_0$. Looking at Eq. \eqref{series}, it is interesting to note that these first two terms are the same for the coherent and quasi-free cases, which only differ starting from the third term. Another interesting fact is that the first three terms do not depend on $p_0$ at all. Heuristically, this means that as the probability distribution centers close to $t=0$ for increasing particle density $r_0$, it also becomes ``less dependent" on the momentum $p_0$. This has implications to the inference problem considered in the next Section.

Finally, in order to give an idea of the multi-time arrival densities, we consider briefly the two-detection probability density. Figure \ref{fig:planewaveB} (b) and (c) shows $p_2 (t_1, t_2)$ for the coherent state (red surfaces)
and the quasi-free state (blue surfaces) for two different values of the particle density $r_0$. By comparing (b) and (c), we observe again that the probability density is getting more and more peaked around $(0,0)$ with increasing $r_0$. Around $(0,0)$ the probability for the coherent state is also larger than the one for the quasi-free state. A thorough study of multi-arrival aspects such as correlations between arrival times is beyond the scope of the present paper, where our main goal is the statistical inference problem described in the next Section.

\section{Statistical inference of single-particle parameters}\label{sec:metrology}

We are now ready to proceed to show how the above
many-particle arrival time processes can be used to infer single-particle properties of the many-particle source states. For concreteness we focus on estimating the mean momentum $p_0=\langle \chi|P|\chi\rangle$ of the particle. Recall that here $\chi$ is the single-particle source state, and we then construct the three types of multi-particle states (Fock, coherent, quasi-free) based on $\chi$. Hence also these multi-particle states depend on $p_0$, and the arrival time distribution \eqref{arrivaltimedistribution} forms a classical statistical model for estimating $p_0$. According to the framework of statistical inference, the inverse of the associated Fisher information quantifies the precision of the optimal estimator, see e.g. \cite{Garthwaite2006}. Here we focus on the Fisher information rather than developing an explicit estimation procedure.

\subsection{General formula of the Fisher information for arrival time measurements}
We first briefly recall the standard general framework of statistical inference, as given in textbooks such as \cite{Garthwaite2006}. For any classical statistical model $p_n({\bf t}, p_0)$ (a probability distribution of the observed data vector ${\bf t}$ of length $n$ given a parameter value $p_0$), the score function $V_{p_0}$ is given by the derivative of the log-likelihood as
$$
V_{n,p_0} = \frac{d}{dp_0}\ln p_n({\bf t},p_0),
$$
and the Fisher information is
$$
I_{n}(p_0) = \mathbb E_n[V_{n,p_0}^2],
$$
where $\mathbb E_n$ denotes the expectation value w.r.t. the distribution $p_n({\bf t},p_0)$.

In our case $p_n({\bf t}, p_0)$ is of course the arrival time distribution of the form \eqref{arrivaltimedistribution}. Here we need to also take into account that the distribution is not necessarily normalised (see Subsection \ref{subsec:noevent}), which means that there is effectively an extra outcome, which we can label as ``NO", for the case where (some of the) $n$ particles fail to arrive. Since the probability of that happening is $1-p_n^{\rm tot}$, we obtain
\begin{align*}
V_{n,p_0}({\bf t}) &= \frac{d}{dp_0} \ln F_n(\Omega(t_n)) +\sum_{k=1}^n \frac{d}{dp_0}\ln \omega(t_k);\\
V_{n,p_0}({\rm NO}) & = -\frac{1}{1-p_n^{\rm tot}}\frac{d}{dp_0} p_n^{\rm tot},
\end{align*}
and the Fisher information reads
\begin{eqnarray*}
I_{n}(p_0) &= \int_0^\infty d^n{\bf t}\, V_{n,p_0}({\bf t})^2 \,F_n(\Omega(t_n))\prod_{i=1}^n \omega(t_i)\\
&+\frac{1}{1-p_n^{\rm tot}}\left(\frac{d}{dp_0} p_n^{\rm tot}\right)^2,
\end{eqnarray*}
where $\int_0^\infty d^n{\bf t} = \int_0^\infty dt_n\int_0^{t_{n-1}}dt_{n-1}\cdots \int_{0}^{t_2}dt_1$ is the integral over the ordered simplex $\{{\bf t}\mid t_1<t_2<\cdots <t_n\}$ of possible arrival time samples. It is useful to realise that the Fisher information can be considerably simplified to give (see Appendix \ref{app:fisher1} for details):
\begin{eqnarray}
\lefteqn{I_{n}(p_0) = \frac{1}{(n-1)!}\int_0^\infty dt \, F_n(\Omega(t))  \omega(t)} & & \nonumber\\ 
&\times&\Bigg[\Omega(t)^{n-1} \left(\frac{d}{dp_0} \ln \left(F_n(\Omega(t)) \omega(t)\right)
\right)^2 \nonumber\\
& & + 2 (n-1) \Omega(t)^{n-2} \frac{d}{dp_0} \ln \left(F_n(\Omega(t))\omega(t)\right) \frac{d \Omega}{dp_0}(t) \nonumber\\
& &+ (n-1)\Omega(t)^{n-2}\int_0^{t}  d\tilde t \, \omega(\tilde t) \left(\frac{d}{dp_0} \ln \omega(\tilde t) \right)^2 \nonumber\\
& &+ (n-1)(n-2)\Omega(t)^{n-3}\left(\int_0^{t}  d\tilde t \, \omega(\tilde t) \frac{d}{dp} \ln \omega(\tilde t) \right)^2\Bigg] \nonumber\\
& &+ \frac{1}{1-p_n^{\rm tot}}\left(\dot \Omega(\infty) \frac{F_{n}(\Omega(\infty))}{(n-1)!} \Omega(\infty)^{n-1}\right)^2,
\label{fisher1}
\end{eqnarray}
where the second and the third term under the outer integral only appear for $n\ge 2$ and the fourth term only for $n \ge 3$. The last term, corresponding to the no-event part, will only appear when $\Omega(\infty)<\infty$.

This formula will be used in what follows
for the numerical calculation of the Fisher information as well as the further analytical result involving sparse beams. We also stress that the formula is not only valid for the mean momentum $p_0$, but works for any single-particle parameter.

\subsection{Numerical examples and discussion}
\begin{figure}[t]
\begin{center}
\includegraphics[scale=0.65]{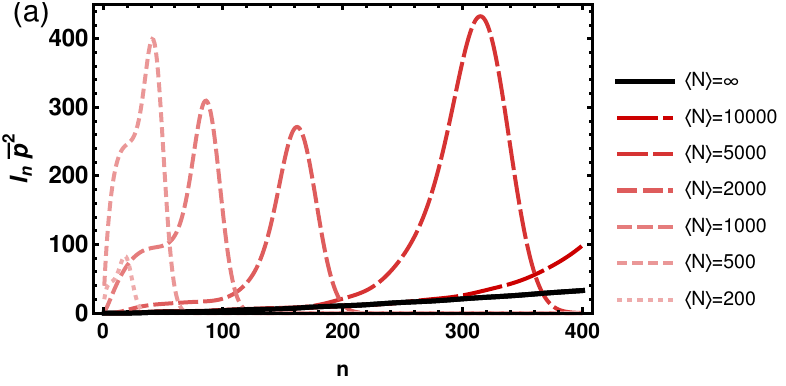}\\
\includegraphics[scale=0.65]{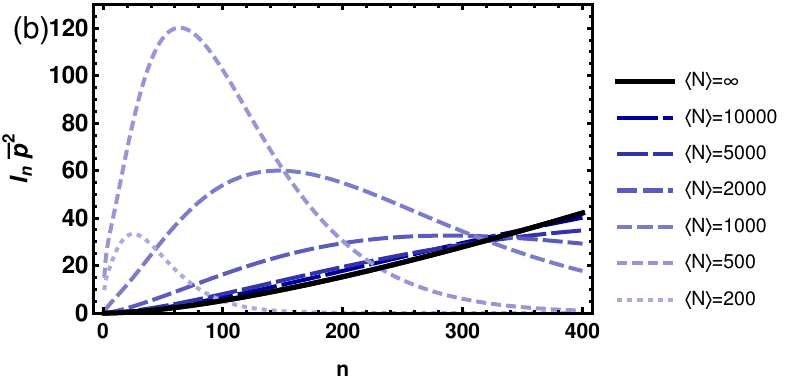}
\end{center}
\caption{\label{fig:fisher} Fisher information $I_n$ for the (a) coherent state and (b) quasi-free state. In each figure, $a \tau/l=0.1$, fixed spatial particle density
$r_0 l=56.42$. The result is shown for different values of the average particle number $\langle N\rangle$; the initial states $\chi_\nnx$ are given by \eqref{initialgaussian} using \eqref{defpnn} and $p_0/\bar p=1$, $x_0/l=-20$, $m l/(\bar p \tau) = 1$.
In addition, the Fisher information based on the analytically computed limit $\omega_{\rm delta}^{\rm beam}(t)$, see \eqref{plane_wave_intensity}, is shown ($N=\infty$, thick, black, solid lines).} 
\vspace{-0.5cm}
\end{figure}

We now examine the Fisher information $I_n (p_0)$, considering the case of the Gaussian single-particle source state \eqref{gaussiandetector} with momentum parameter $p_0$, detected by a delta-detector as described in the preceding section. We set the true value of the parameter as $p_0/\bar p = 1$; the corresponding Fisher information is shown for coherent states (Fig. \ref{fig:fisher}(a)) as well as quasi-free states  (Fig. \ref{fig:fisher}(b)).

From Fig. \ref{fig:fisher}(a) and (b), we see that the shape of the Fisher information graph as a function of the number of detections, $n$, crucially depends on the corresponding state (coherent or quasi-free). However, in both cases, the Fisher information clearly has a maximum, implying that there is an optimal number of detections. This maximum moves to larger number of detections $n$ when the average particle number $\langle N \rangle$ increases.

\begin{figure}[t]
\begin{center}
\includegraphics[scale=0.65]{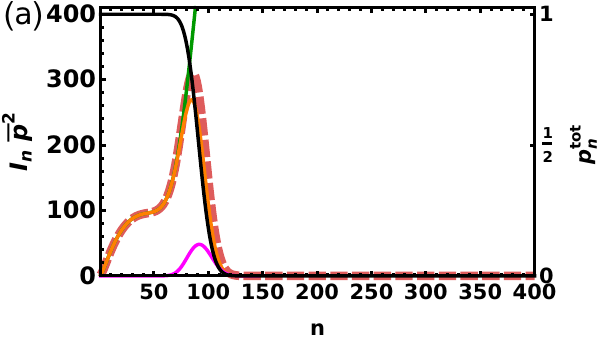}\\
\includegraphics[scale=0.65]{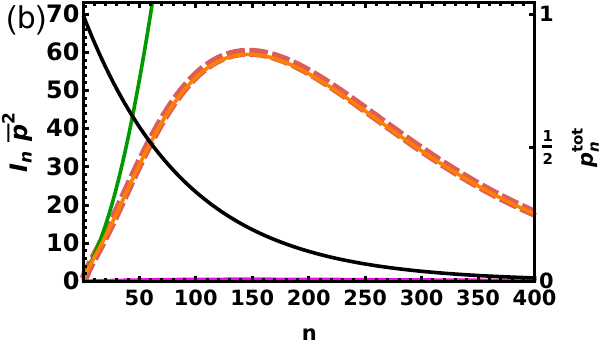}
\end{center}
\caption{\label{fig:fisherdetails} Decomposition of the Fisher information for the (a) coherent state and (b) quasi-free state, for $\langle N \rangle = 1000$. Fisher information $I_n(p_0)$ (thick, dashed lines), conditional Fisher information $I_n^{(c)}(p_0)$ (green lines), approximated Fisher information $I_n^{(a)}(p_0)$ (orange lines), the no-detection contribution (magenta lines), and the total detection probability $p^{\rm tot}_n$ (black line, right axis). Other parameters are as in Fig. \ref{fig:fisher}.} 
\vspace{-0.5cm}
\end{figure}

To understand this observation, let us introduce the conditional Fisher information $I_n^{(c)}(p_0)$ under the condition that at least $n$ detections have happened; this is the Fisher information for the probability distribution $p_n^{(c)} ({\bf t}) = p_n ({\bf t})/p_n^{\rm tot}$, renormalised to have total probability one. One can easily show that the actual Fisher information and the conditional Fisher information are related by
\begin{eqnarray}
I_n(p_0) = p_n^{\rm tot}I_n^{(c)}(p_0) + \frac{1-p_n^{\rm tot}}{p_n^{\rm tot}}   \left(\frac{d}{dp_0} \ln (1-p_n^{\rm tot})\right)^2\,.\nonumber\\
\end{eqnarray}
Now, the maximum of $I_n(p_0)$ arises because, on the one hand, the probability $p_n^{\rm tot}$ of having $n$ detections decreases to zero as $n \to \infty$ (as shown before, see \eqref{pntotlimit}), but on the other hand, the Fisher information $I_n^{(c)}(p_0)$ conditioned on $n$ detections is increasing as $n \to \infty$, see Fig. \ref{fig:fisherdetails}.

In the case of the coherent state, the probability $p_n^{\rm tot}$ decreased relatively abruptly after a certain threshold value. As it can be seen in the Fig. \ref{fig:fisherdetails}, this threshold value approximates quite well the position of the maximum of the Fisher information. We emphasise that the threshold is significantly different from the average particle number $\langle N \rangle$, which is much larger. The reason is that also the properties of the detector (not only the source state) play a role here.

For the quasi-free state the probability $p_n^{\rm tot}$ does not show a similar clearly discernible threshold value, but the above reasoning concerning the appearance of the maximum apply in this case as well.

The beam limit $\langle N \rangle = \infty$ is also shown in Fig. \ref{fig:fisher}, for both coherent and a quasi-free state; the convergence to the limit case as $\langle N \rangle \to \infty$ is clearly visible. Furthermore (as obtained in the preceding section) $p_n^{\rm tot}$ is always one in the beam case, and therefore one would expect the Fisher information to be increasing with $n$, in contrast to the finite-particle case discussed above. This is indeed clearly shown in the Fig. \ref{fig:fisher}.

\subsection{Different particle densities and the limit of sparse beams}

We now proceed to study how information on $p_0$ (obtained from the arrival time data) depends on the characteristics of the beam source. Recall that in addition to the single-particle parameter $p_0$, the arrival time distributions in the beam case are characterised (only) by the spatial particle density $r_0$, the detection strength $a$, and the particle mass $m$. It is therefore particularly interesting to study how the Fisher information of $p_0$ depends on these variables. Here we consider only the beam characteristic $r_0$, leaving the study of the other variables for a future work.

\begin{figure}[t!]
\begin{center}
\includegraphics[scale=0.85]{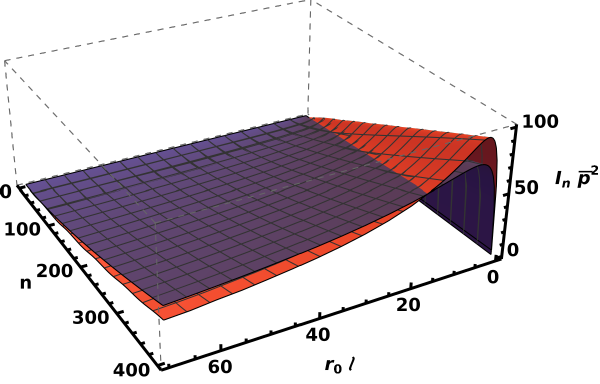}
\end{center}
\caption{\label{fig:fisher2} Fisher information $I_n(p_0)$ for the coherent state (red surface) and quasi-free state (blue surface) in the beam limit $\langle N \rangle=\infty$ 
for different spatial particle densities $r_0$. For $r_0=0$ the obtained analytical results \eqref{analc} and \eqref{analq} are used.
Other values are as in Fig. \ref{fig:fisher}.}
\vspace{-0.5cm}
\end{figure}

In order to get a rough idea of the effect of $r_0$, we consider in Fig. \ref{fig:fisher2} the Fisher information in the beam case for varying $r_0$, for the different types of states, coherent (red surface) and quasi-free (blue surface). Interestingly, it turns out that the extreme cases (large and small $r_0$) are analytically tractable.

First of all, we observe that the Fisher information is decreasing with increasing particle density $r_0$. The heuristic reason for this is that for increasing $r_0$ the arrival probability density is expected to peak more and more around zero; see Fig.  \ref{fig:planewaveB}, the series expansion \eqref{firstarrivalexpansion}, and the associated discussion. Note, in particular, that the value $p_1(0)=ar_0$, the first detection probability density at zero, contains no information on $p_0$ at all, so the closer the arrival is to zero, the less information on $p_0$ it provides. In line with this heuristic reasoning, we can in fact rigorously prove (see Appendix \ref{app:fisher1}) that $\lim_{r_0\rightarrow\infty}I_n(p_0) =0$ for both coherent and quasi-free cases, that is, the information tends to zero as the particle density increases to infinity, for any fixed number $n$ of detections. Therefore, this ``dense beam" limit does not yield any information on the momentum parameter $p_0$.

Next, we observe from Fig. \ref{fig:fisher2} that the Fisher information peaks at some $r_{max}$ for each $n$, so it is increasing in $r_0$ for small values of $r_0$. This leads us to consider the other extreme case $r_0\approx 0$ (sparse beam), the obvious question being whether the information is zero also at this limit. Surprisingly, even though the arrival time distribution escapes to infinity in this limit (we need to wait infinitely long to get detections), the Fisher information converges to a \emph{nonzero} limit, which also has a very simple analytical form. In order to state the result, we first denote
\begin{eqnarray}
I_\infty(p_0) := \lim_{t\to\infty}\left(\frac{\frac{d}{dp}\omega_{\rm delta}^{\rm beam}(t)}{\omega_{\rm beam}^{\rm delta}(t)}\right)^2 = \frac{a^2 m^2}{p_0^2 \left(p_0 + \frac{a m}{2}\right)^2},
\end{eqnarray}
where the first equality is the definition while the second is a consequence of \eqref{omegaapprox} and \eqref{domegaapprox}. Now, in Appendix \ref{app:fisher1} we show (with a fully rigorous proof) that for a coherent state,
\begin{eqnarray}
\lim_{r_0\to 0} I_n(p_0) = n \, I_\infty(p_0)
\label{analc}
\end{eqnarray}
and for a quasi-free state,
\begin{eqnarray}
\lim_{r_0\to 0} I_n(p_0) = \frac{n}{n+2} \, I_\infty(p_0)\, .
\label{analq}
\end{eqnarray}
These results show a striking qualitative difference between the two types of states: the Fisher information increases linearly with the sample size for the coherent case, but saturates to the maximum value $I_\infty(p_0)$ for the quasi-free case. Therefore, the benefit of multiple detections is smaller in the latter case.

We note that for the numerical values used in the Fig. \ref{fig:fisher2} we get $I_\infty(p_0) = 0.009 /\bar p^2$, which is nonzero but small enough not to be visible with the resolution of the figure.

In order to intuitively understand the limit $r_0\rightarrow 0$, we also show in Appendix \ref{app:fisher1} that the limit Fisher information actually coincides with the one obtained from a (hypothetical) \emph{time-stationary} model where the intensity $\omega(t)$ is a constant function of time. This is not a trivial result as the arrival time distributions themselves do \emph{not} converge as $r_0\rightarrow 0$. However, it can be understood heuristically by noting that the distribution ``escapes to infinity'' with vanishing particle density, so it takes longer for the particles to arrive, and hence, for any fixed large time $T$, the detection probability distributions will be essentially supported in times $t>T$ for small enough $r_0$. Therefore, only the value of $\omega_{\rm beam}^{\rm delta}(t)$ at large $t$ plays a role, and this is approximately equal to the time-independent constant given by \eqref{constantomega}. Hence, by scaling the time by the density $r_0$ we might expect to obtain a time-stationary model. However, as we are taking two limits here ($r_0\rightarrow 0$ and $t\rightarrow\infty$), the above argument is only heuristic, and so the result requires the mathematically rigorous argument based on the dominated convergence theorem given in Appendix \ref{app:fisher1}.

\section{Conclusion and outlook}\label{sec:outlook}

In this paper, 
we have presented a general description of a many-particle arrival time detection process based on a phenomenological absorption model for the detector.
We have first derived the resulting arrival time distributions in the case of Fock, coherent, and quasi-free source states, each with finite average particle number, see \eqref{arrivaltimedistribution}, where the intensity function is given by \eqref{intensity}. We have shown that even the distribution of the first arrival time depends significantly on the type of the many-particle state. In particular, it is very different from the distribution we obtain if the system only consists of one particle.

We have then considered the limit of a delta detector located at a single point in space, with the resulting intensity function \eqref{delta_intensity}. Subsequently we pass further to the limit of infinite number of particles forming a spatially uniform beam with a single momentum value.
In that case, the intensity function has the explicit form \eqref{plane_wave_intensity}.

After the derivation of the arrival time distributions, we have proceeded to consider them as statistical models; this allowed us to quantify the information on the single-particle parameters using the classical Fisher information. Our main focus here was on a detailed study of the case where the arrival time sample originates from a spatially uniform beam with a definite momentum detected by the point detector model. We derived an analytical formula for the associated Fisher information \eqref{fisher1}, comparing the result numerically to the case where the source state has finite average number of particles. Interestingly, for the finite particle case the Fisher information has a maximum for a certain value of $n$, the number of detections, but increases in $n$ for the beam source state.

We stress here that in the case of the spatially uniform beam, the mechanism by which information on the sole single-particle parameter (momentum) becomes encoded in the arrival time data is well isolated from other effects; in particular, the position distribution of the particles contains no information on the parameter. Hence this case is particularly instructive and tractable not only analytically but also in terms of physical intuition. Even more surprisingly, the corresponding \emph{inverse problem} of statistical inference, which in general is extremely complex due to the correlated sample, becomes tractable as well. In the further limit of sparse beams, we have even obtained fully explicit analytical formulas for the Fisher information, in terms of the number of detections $n$, the momentum parameter, as well as the two remaining auxiliary variables (detection strength and the mass of the particle), see \eqref{analc}, \eqref{analq}. These formulas also show an interesting qualitative difference between the coherent and quasi-free cases, which is not apparent in the more generic situation.

Therefore, our explicit case of momentum parameter inference provides an important starting point for understanding how single-particle information is generally reflected in temporal data structures obtained by observing beams of particles. This lays the foundations for extensions in future work on many-particle arrival time processes; for instance, one can consider source states with varying quantum correlations, or study the effect of particle-particle interactions using multi-parameter inference. Finally, our results contribute to the fundamental understanding of temporal data arising from quantum systems consisting of freely evolving particles, as opposed to constrained information carriers.

\begin{acknowledgements}
A.R acknowledges that this publication has emanated from research
funded by Taighde \'Eireann – Research Ireland
under Grant Number 19/FFP/6951. 
\end{acknowledgements}

\appendix

\section{Calculating the no-event probability in Subsection \ref{subsec:noevent}}\label{app:noevent}

We first note that in the coherent and quasi-free cases the function $F$ defining the arrival time distributions \eqref{arrivaltimedistribution} is a monotone decreasing nonnegative function real analytic over $\Omega>0$. (In the Fock case the point $\Omega =N$ breaks the analyticity). Furthermore, in each case $F$ satisfies $F(0)=1$, $F(\infty)=0$, and the function $F_n=(-1)^nF^{(n)}$ is nonnegative, satisfying the decay condition 
\begin{equation}\label{decay}
\lim_{\Omega\rightarrow\infty}\Omega^n F_n(\Omega)=0, \quad n=0,1,2\ldots.
\end{equation}
Now, in the main text we obtained the total detection probability for $n$ detections, see eq. \eqref{totalprob}.
Using integration by parts with $F_n=(-1)^n F^{(n)}$ we therefore obtain the recurrence relation
\begin{equation}\label{recur}
p_n^{\rm tot} = p_{n-1}^{\rm tot} - \frac{F_{n-1}(\Omega(\infty))}{(n-1)!} \Omega(\infty)^{n-1},
\end{equation}
where the initial case is $p_1^{\rm tot}=-\int_0^{\Omega(\infty)} F^{(1)}(u)du=F(0)-F(\Omega(\infty))=1-F(\Omega(\infty))$. There are now two qualitatively different cases, depending on whether $\Omega(\infty)$ is finite or not. Assume first that $\Omega(\infty)=\infty$. Then, \eqref{decay} and \eqref{recur} immediately imply that $$p_n^{\rm tot}=p_{n-1}^{\rm tot}=p_1^{\rm tot}=1-F(\infty)=1$$ for all $n$. Assume then that $\Omega(\infty)<\infty$. Then the recurrence \eqref{recur} implies (due to positivity of $F_n$) that $p_n^{\rm tot}$ is decreasing in $n$. Furthermore, we can expand the recurrence as
\begin{align*}
p_n^{\rm tot} &=p_1^{\rm tot} -\sum_{k=1}^{n-1} (-1)^k \frac{F^{(k)}(\Omega(\infty))}{k!} \Omega(\infty)^k\\
&= 1- F(\Omega(\infty)) -\sum_{k=1}^{n-1}\frac{F^{(k)}(\Omega(\infty))}{k!} (-\Omega(\infty))^k\\
&= 1- \sum_{k=0}^{n-1}\frac{F^{(k)}(\Omega(\infty))}{k!} (0-\Omega(\infty))^k.
\end{align*}
But the sum in the last line is a partial sum of the Taylor series of the function $F$ around the point $\Omega(\infty)$ evaluated at $0$. If $F$ is analytic (as is the case for coherent and quasi-free states), the sum converges to $F(0)$ and we obtain 
$$
\lim_{n\rightarrow \infty} p_n^{\rm tot} = 1- F(0)=0,
$$
showing that the total detection probability approaches zero at the limit $n\rightarrow \infty$. The same holds trivially in the Fock case because $F^{(n)}$ is identically zero for
 $n>N$.

\section{Convergence to the delta detector equation\label{app:deltaequation}}

Here we consider in detail the convergence of the intensity function at the delta-detector limit, and then derive in detail the analytical form of the resulting intensity function.

First, we recall from the main text that the intensity function is given by $\omega_\epsilon(t) := \langle N\rangle \gamma_\epsilon |\langle \phi_\epsilon|\chi_t\rangle|^2$, where $$\delta_\epsilon(x):=\sqrt{\gamma_\epsilon/a}\,\phi_\epsilon(x)\rightarrow \delta(x).$$ 
To make a rigorous argument, we specify in detail what this means: we assume that $s_0:=\sup_{\epsilon>0} \int_\mathbb R |\delta_\epsilon(x)|dx<\infty$, and for any bounded function $f:\mathbb R\to \mathbb C$ continuous (at least) at $x=0$, we have
\begin{align}\label{deltalimit}
\lim_{\epsilon\rightarrow 0} \int_{\mathbb R} f(x)\delta_{\epsilon}(x) dx = f(0).
\end{align}
Note that if each $\delta_\epsilon(x)$ is positive, then this follows simply if $\int_{\mathbb R} \delta_\epsilon(x)dx=1$ and $\lim_{\epsilon \rightarrow 0}\int_{x:|x|>r} \delta_\epsilon(x)dx=0$ for any $r>0$. This clearly holds for our choice \eqref{gaussiandetector} in the main text.

\emph{We will show that once a family $\phi_\epsilon(t)$ is chosen as above, there exists a sequence $(\epsilon_k)_{k\in \mathbb N}$ with $\lim_{n\rightarrow \infty} \epsilon_n=0$, such that $$\lim_{n\rightarrow \infty} \omega_{\epsilon_n}(t)=\omega_{\rm delta}(t)$$
for almost all $t\geq 0$.}

To proceed, recall that for each $\epsilon$ we have the integral equation
\begin{equation}\label{inteqappendix}
f_\epsilon(t) = h_{0,\epsilon}(t) - \frac 12\int_0^t g_{\epsilon}(t-s)f_{\epsilon}(s)\,ds,
\end{equation}
where
\begin{align*}
f_\epsilon(t) &= \sqrt{\gamma_\epsilon/a} \langle \phi_\epsilon|\chi_t\rangle, \\
h_{0,\epsilon} &= \sqrt{\gamma_\epsilon/a} \langle \phi_\epsilon|e^{-it H/\hbar}\chi\rangle, & g_\epsilon(t) &= \gamma_\epsilon\langle \phi_\epsilon|e^{-itH/\hbar}\phi_\epsilon \rangle,
\end{align*}
and we have assumed the source state $\chi(x)$ integrable as stated in the main text. We can now write
\begin{align*}
h_{0,\epsilon}(t) &= \int dx\,\overline{\delta_\epsilon(x)} \left[\int K_t(x,x')\chi(x')dx' \right],\\
g_\epsilon(t) & = a \int\int dx dx'\, \overline{\delta_\epsilon(x)} K_t(x,x') \delta_{\epsilon}(x'),
\end{align*}
where $K_t(x,x') = \sqrt{m/(2\pi i\hbar t)} \exp[-m(x-x')^2/(2i\hbar t)]$ is the free propagator. Now $|K_t(x,x')|\leq k_0 t^{-\frac 12}$ where $k_0=\sqrt{m/(2\pi\hbar)}$, so $K_t(x,x')$ is a bounded continuous function of $x,x'$ for each fixed $t$, hence $|\int K(x,x') \chi(x')dx'|\leq k_0t^{-\frac 12}\int|\chi(x')|dx'$ for all $x$. Therefore, \eqref{deltalimit} applies, and we get $$\lim_{\epsilon\rightarrow 0}h_{0,\epsilon}(t) = \int K(0,x')\chi(x') dx'= f_0(t)$$ for each $t>0$, where $f_0(t) = [e^{-itH/\hbar}\chi](0)$ as in the main text. Furthermore,
$$
|h_{0,\epsilon}(t)|\leq k_0\,s_0\,t^{-\frac 12}\int|\chi(x)|dx<\infty,
$$
and $|f_0(t)|\leq k_0t^{-\frac 12}\int |\chi(x)| dx$; since these bounds are integrable over $[0,t_0)$ for any $t_0>0$, we have $$\lim_{\epsilon\rightarrow 0}\int_0^{t_0} |h_{0,\epsilon}(t)-f_0(t)|dt=0$$ by the dominated convergence theorem. Therefore, $(h_{0,\epsilon})$ converges to $f_0$ in $L^1[0,t_0)$ for all $t_0>0$, that is, locally in $L^1$.

Furthermore, the bound $|K_t(x,x')|\leq k_0 t^{-\frac 12}$ also shows that $|g_\epsilon(t)|\leq a k_0 s_0^2 t^{-\frac 12}$, so using the continuity of $x\mapsto K_t(x,x')$ and $x'\mapsto K_t(x,x')$ we get first the pointwise limit
$$
\lim_{\epsilon\rightarrow 0}g_\epsilon(t) = a K_t(0,0) = a\sqrt{\frac{m}{2\pi i\hbar t}},
$$
and then local $L^1$-convergence follows again by dominated convergence.

Now, the local $L^1$-convergence implies (see \cite{Lew1972}) that the solution to \eqref{inteqappendix}, that is, the function $f_\epsilon(t)$, is determined by $h_{0,\epsilon}(t)$ and $g_\epsilon(t)$ for each $\epsilon$, and converges to the solution $f(t)$ of the delta-detector equation \eqref{renewal}, again in the sense of local $L^1$-convergence. This implies that some subsequence converges almost everywhere, which gives the above claim.

Now the solution of \eqref{renewal} is uniquely determined by the initial function $f_{\rm free}(t)$, and it can be found explicitly, using e.g. the pseudo-inverse approach of \cite{Lew1972}, or a Laplace transform method \cite{Bellman1984}. According to the latter, the solution is of the form
\begin{align}
f(t) &= g(0)f_{\rm free}(t) + (f_{\rm free} *g')(t),\label{sol1}\\
&=f_{\rm free}(0)g(t) + (f_{\rm free}'*g)(t)\label{sol2},
\end{align}
where prime denotes derivative, $(f_1*f_2)(t) = \int_0^t f_1(t-s)f_2(s)ds$ is the Laplace convolution, and the function $g(t)$ is determined by the equation
$$
g(t) = 1 -\frac{d}{\sqrt \pi}\int_0^t \frac{g(s)}{\sqrt{t-s}}.
$$
Letting $G(s)$ denote the Laplace transform of $g(t)$ and using the Laplace transform $\sqrt{\pi/s}$ of $1/\sqrt t$, the convolution theorem gives $G(s)= 1/s - a s^{-\frac 12}G(s)$, hence
$$
G(s) = \frac{1}{s(1+ a s^{-\frac 12})} = \frac{1}{\sqrt{s}(\sqrt{s}+a)},
$$
which (by e.g. \cite{Abramovitz1965}) is the Laplace transform of
\begin{equation}\label{kern}
g(t) = e^{a^2t} {\rm erfc}(a \sqrt t).
\end{equation}

Note that the dependence on the source state $\chi$ is entirely contained in $f_{\rm free}(t)=(e^{-itH/\hbar}\chi)(0)$, which determines $f(t)$ through direct integration in \eqref{sol1} or \eqref{sol2}. Since \eqref{sol1} is linear in $f_{\rm free}(t)$, we can express the solution for an arbitrary $\chi$ by expanding $f_{\rm free}(t)$ in a suitable basis. In particular, using the momentum representation for $\chi$ we have
$$
f_{\rm free}(t) = \frac{1}{\sqrt{2\pi\hbar}}\int_{-\infty}^\infty e^{-it\frac{p^2}{2m\hbar}}\hat\chi(p)dp, 
$$
so we only need the solution for the initial function $f_{\rm free}(t)=\frac{1}{\sqrt{2\pi\hbar}} e^{-it p^2/(2m\hbar)}$. By direct substitution we obtain the functions $f_p(t)$ given in \eqref{deltapsol} in the main text.

\section{Spatial particle density\label{app:spatial}}

Here we derive the form \eqref{particledensity}. Let $Q_I$ be as \eqref{sec:beam} (that is, the spectral projection of the standard position operator for the interval $I$). A measurement of $Q_I$, then, yields 1 if the particle lies in $I$ and zero otherwise, and therefore the 2nd quantised observable $q(Q_I)$ (acting on the many-particle level) represents the number of particles in $I$. Now, let $N(I)$ denote the number of particles in $I$ given that the state of the many-particle system is $\rho$. Then the distribution of the random variable $N(I)$ is (in each of the three cases) given by the characteristic function
\begin{align*}
C_{N(I)}(t) & = {\rm tr}[\rho e^{it q(Q_I)}] =\sum_{N=0}^\infty p_N \langle \chi|e^{it Q_I}\chi\rangle^N\\
&=\sum_{N=0}^\infty p_N \left(\int_I e^{it}|\langle \chi|x\rangle|^2+ \int_{\mathbb R\setminus I} |\langle \chi|x\rangle|^2\right)^N\\
&= \sum_{N=0}^\infty p_N \left(1+ (e^{it}-1)\int_I |\langle\chi|x\rangle|^2dx\right)^N,
\end{align*}
where $(p_N)$ is the particle number distribution (which is different in each of the three cases). Therefore the mean particle number in the interval $I$ is
$$\langle N(I)\rangle=-i \frac{d}{dt}C_{N(I)}(t)|_{t=0} = \langle N\rangle \int_I |\langle\chi|x\rangle|^2dx,$$
and hence $r(x):=\langle N\rangle |\langle\chi|x\rangle|^2$ is indeed the particle number density as claimed in the main text.

We now consider the particle number distribution in the plane wave beam limit in the three relevant cases, showing that it indeed converges to a limit which does not contain any single-particle information.
 
Starting with the Fock states, the spatial particle number distribution is given by
\begin{equation}\label{charnum}
C_{N(I)}(t) =\left(1+ (e^{it}-1)\frac{1}{N}\int_I r(x)\right)^N.
\end{equation}
In the beam limit we have $r(x)\rightarrow r_0$ for all $x$ (as described in the main text), and $N=\langle N\rangle\rightarrow\infty$ so we get $C_{N(I)}(t)\rightarrow e^{r_0|I|(e^{it}-1)}$ (independently on how the states \eqref{planewave1} are chosen). This is just a Poisson distribution with mean $r_0|I|$, showing that the particles are scattered randomly in space with average rate of $r_0$ per unit length.

In the case of coherent sources, the spatial number distribution is Poisson even without taking the plane wave limit:
\begin{equation}\label{charcoh}
C_{N(I)}(t) = e^{(e^{it}-1)\int_I r(x) dx}.
\end{equation}
Hence the limit yields $C_{N(I)}(t)\rightarrow e^{r_0|I|(e^{it}-1)}$, that is, exactly the same as obtained by starting with states of fixed particle number.

Finally, in the case of quasi-free source the spatial number distribution is obtained by summing a geometric series to get
\begin{equation}\label{charq}
C_{N(I)}(t) = \left(1-(e^{it}-1)\int_I r(x)dx\right)^{-1},
\end{equation}
and the plane wave limit yields $$C_{N(I)}(t) \rightarrow (1- r_0|I|(e^{it}-1))^{-1},$$ which is a Geometric distribution. In each case the only remaining parameter is the constant spatial density $r_0$.

\begin{widetext}
\section{Fisher information for arrival time measurement\label{app:fisher1}}

Fix $n\in \mathbb N$. In the main text we obtained the arrival time distributions in the form
$$
p_n(\mathbf t) = F_n(\Omega(t_n)) \prod_{k=1}^n \omega(t_k), \quad p_n({\rm NOT}) = 1-p_{n}^{\rm tot}.
$$

Here we note that 
$$
\frac{d}{d\Omega} F_n(\Omega) = -F_{n+1}(\Omega)
$$
for all $n=0,1,\ldots$. The unknown parameter to be estimated is in the main text the momentum $p_0$, but the following derivation holds for any parameter, which we denote simply by $p$. We assume that $\omega(t)$ depends on $p$, and hence the  distribution $p_n({\bf t})$ becomes a parametric statistical model \cite{Garthwaite2006}. We then have the score function for the unknown parameter $p$ as
\begin{align*}
V_p({\bf t}) &:= \frac{d}{dp} \ln p_n({\bf t}) = \frac{d}{dp}\left(\ln F_n(\Omega(t_n))\omega(t_n))+\sum_{k=1}^{n-1}\ln\omega(t_k)\right), & V_p({\rm NOT}) =\frac{d}{dp} \ln (1-p_n^{\rm tot}), 
\end{align*}
where we have separated the part only depending on $t_n$. The Fisher information from $n$ detections is
$$
I_n(p) :=\mathbb E[V_p] = I_{n,0}(p) + I_n({\rm NOT}),
$$
where the second term 
$$
I_n({\rm NOT})= \frac{1}{1-p_n^{\rm tot}}\left(\frac{d}{dp}p_n^{\rm tot}\right)^2
$$
corresponds to the no-event part, and we focus now on simplifying the expression for the first term
\begin{align*}
I_{n,0}(p) &:= \int_{0}^\infty dt_n F_n(\Omega(t_n))\omega(t_n) \int_0^{t_n} \omega(t_{n-1})dt_{n-1}\cdots \int_{0}^{t_2} dt_1\,\omega(t_1) \,\left(\frac{d}{dp} \ln p_n({\bf t})\right)^2.
\end{align*}
In order to simplify the calculations, we make the substitution $u=\Omega(t)$ (so that $du = \omega(t) dt$) to each of the integrals; here we need to assume that $\omega(t)\neq 0$ for all $t>0$. Denoting
\begin{align*}
h(u) &:= \frac{d}{dp}[\ln F_n(\Omega(t))\omega(t)]_{t=\Omega^{-1}(u)} = -\frac{F_{n+1}(u)}{F_{n}(u)}\dot \Omega(\Omega^{-1}(u)) + v(u), & v(u) &:= \frac{d}{dp}[\ln \omega(t)]_{t=\Omega^{-1}(u)} = \frac{\dot\omega(\Omega^{-1}(u))}{\omega(\Omega^{-1}(u))},
\end{align*}
where $\dot\omega(t) = \frac{\partial}{\partial p}\omega(t)$, and noting that $\Omega(0)=0$ we get
\begin{align*}
I_{n,0}(p) &= \int_{0}^{\Omega(\infty)} du\, F_n(u)\, \int_0^{u} du_{n-1}\cdots \int_{0}^{u_2} du_1 \,\left(h(u)+\sum_{k=1}^{n-1}v(u_k)\right)^2\\
&= \int_{0}^{\Omega(\infty)} du\, F_n(u)\, \int_0^{u} du_{n-1}\cdots \int_{0}^{u_2} du_1 \,\left(h(u)^2+2h(u)\sum_{k=1}^{n-1}v(u_k) +\sum_{k=1}^{n-1}v(u_k)^2+ 2\sum_{k=2}^{n-1}\sum_{m=1}^{k-1} v(u_k)v(u_m)\right),
\end{align*}
with the understanding that the second and the third term are only present for $n\geq 2$, and the last one only for $n\geq 3$. In order to simplify the iterated integrals we use the Cauchy formula for repeated integration:
\begin{equation}\label{integralformula}
\int_0^{u} du_{k-1}\int_0^{k-1}\cdots \int_0^{u_3} du_2\int_{0}^{u_2} du_1 f(u_1) = \frac{1}{(k-2)!}\int_0^{u} (u-u')^{k-2} f(u')du',
\end{equation}
which holds for any function $f(u)$ and all $k=2,3,\ldots$, and is easy to prove by induction.
We note the special case with $f(u)=1$ (constant), which gives
\begin{equation}\label{simpleformula}
\int_0^{u} du_{k-1}\int_0^{k-1}\cdots \int_0^{u_3} du_2\int_{0}^{u_2} du_1= \frac{1}{(k-2)!}\int_0^{u} (u-u')^{k-2} du'= \frac{u^{k-1}}{(k-1)!}.
\end{equation}
We can now write
$$
I_{n,0}(p) = \int_0^{\Omega(\infty)} du F_n(u) \left(h(u)^2 q^{(1)}_n(u) + 2h(u) q^{(2)}_n(u) + q^{(3)}_n(u) + 2 q^{(4)}_n(u)\right),
$$
with
\begin{align*}
q^{(1)}_n(u) & = \int_0^{u} du_{n-1}\cdots \int_{0}^{u_2} du_1 = \frac{1}{(n-1)!} u^{n-1}, \quad n\geq 1,
\end{align*}
where we have used \eqref{simpleformula};
\begin{align*}
q^{(2)}_n(u)&=\sum_{k=1}^{n-1} \frac{1}{(k-1)!}\int_0^{u} du_{n-1}\cdots \int_{0}^{u_{k+1}} du_k v(u_k) u_k^{k-1}\\
&= \sum_{k=1}^{n-1} \frac{1}{(k-1)!} \frac{1}{(n-k-1)!}\int_0^u du'(u-u')^{n-k-1}v(u')u'^{k-1},\\
&= \frac{1}{(n-2)!}\int_0^u du' \left[\sum_{k=1}^{n-1}\binom{n-2}{k-1}u'^{k-1}(u-u')^{n-k-1}\right]v(u'),\\
&= \frac{1}{(n-2)!}\int_0^u du' \left[\sum_{k=0}^{n-2}\binom{n-2}{k}u'^{k}(u-u')^{n-2-k}\right]v(u'),\\
&= \frac{1}{(n-2)!}\int_0^u du' (u'+u-u')^{n-2} v(u') = \frac{1}{(n-2)!}u^{n-2}\int_0^u du' v(u'),\quad n\geq 2,
\end{align*}
where we have used first \eqref{simpleformula}, then \eqref{integralformula} (applied to the function $f(u)=u^{k-1}v(u)$), and the Binomial theorem;
\begin{align*}
q^{(3)}_n(u)&:=\sum_{k=1}^{n-1} \frac{1}{(k-1)!}\int_0^{u} du_{n-1}\cdots \int_{0}^{u_{k+1}} du_k v(u_k)^2 u_k^{k-1}= \frac{1}{(n-2)!} u^{n-2}\int_0^u du' v(u')^2, \quad n\geq 2,
\end{align*}
where we have proceeded as with $q^{(2)}_n(u)$, and
\begin{align*}
q^{(4)}_n(u) &=\sum_{k=2}^{n-1} \int_0^{u} du_{n-1}\cdots \int_0^{u_{k+1}}du_k v(u_k) \sum_{m=1}^{k-1}\frac{1}{(m-1)!}\int_0^{u_{k}}du_{k-1}\cdots \int_{0}^{u_{m+1}} du_m v(u_m) u_m^{m-1}\\
&=\sum_{k=2}^{n-1} \int_0^{u} du_{n-1}\cdots \int_0^{u_{k+1}}du_k v(u_k) q_k^{(2)}(u_k)\\
&=\sum_{k=2}^{n-1} \frac{1}{(k-2)!}\int_0^{u} du_{n-1}\cdots \int_0^{u_{k+1}}du_k \,v(u_k) u_k^{k-2}\int_0^{u_k} du' v(u')\\
&=\sum_{k=2}^{n-1} \frac{1}{(k-2)!}\frac{1}{(n-k-1)!} \int_0^u du'' (u-u'')^{n-k-1}v(u'') u''^{k-2}\int_0^{u''} du' v(u')\\
&=\frac{1}{(n-3)!}\int_0^u du''\left[\sum_{k=2}^{n-1} \binom{n-3}{k-2}  u''^{k-2}(u-u'')^{n-k-1} \right]v(u'') \int_0^{u''} du' v(u')\\
&=\frac{1}{(n-3)!}\int_0^u du''\left[\sum_{k=0}^{n-3} \binom{n-3}{k}  u''^{k}(u-u'')^{n-3-k} \right]v(u'') \int_0^{u''} du' v(u')\\
&=\frac{1}{(n-3)!}\int_0^u du''(u''+u-u'')^{n-3}v(u'') \int_0^{u''} du' v(u')\\
& =\frac{1}{(n-3)!}u^{n-3}\int_0^u du''v(u'') \int_0^{u''} du' v(u')=\frac{1}{2(n-3)!}u^{n-3}\left(\int_0^u du'v(u')\right)^2, \quad n\geq 3,
\end{align*}
where we have used the previous result for $q^{(2)}_k(u_k)$, then again \eqref{integralformula} (applied to the function $f(u) = v(u) u^{k-2}\int_0^{u} du' v(u')$), and the Binomial theorem.

Now we can back-substitute the original time variable $t = \Omega^{-1}(u)$ to get
\begin{align*}
q^{(1)}_n(\Omega(t)) & = \frac{1}{(n-1)!} \Omega(t)^{n-1}, \\
q^{(2)}_n(\Omega(t)) & = \frac{1}{(n-2)!} \Omega(t)^{n-2}\int_0^t dt' \dot\omega(t'), & q^{(3)}_n(\Omega(t))&=\frac{1}{(n-2)!}\Omega(t)^{n-2}\int_0^t dt'\frac{\dot \omega(t')^2}{\omega(t')}\\
q_n^{(4)}(\Omega(t)) &= \frac{1}{(n-3)!} \Omega(t)^{n-3}\int_0^t dt'\dot \omega(t') \int_0^{t'} dt''\dot\omega(t''),
\end{align*}
and hence the Fisher information is
\begin{align*}
I_{n,0}(p) &= \frac{1}{(n-1)!}\int_0^{\infty} dt F_n(\Omega(t))\omega(t) \Omega(t)^{n-1}h(\Omega(t))^2 \\
& + \frac{1}{(n-2)!}\int_0^{\infty} dt F_n(\Omega(t))\omega(t) \Omega(t)^{n-2} \left( 2h(\Omega(t))\int_0^t dt' \dot\omega(t') + \int_0^t dt'\frac{\dot \omega(t')^2}{\omega(t')}\right) \\
&+ \frac{1}{(n-3)!} \int_0^{\infty} dt F_n(\Omega(t))\omega(t)\Omega(t)^{n-3}\left(\int_0^t dt'\dot \omega(t')\right)^2.
\end{align*}
which can be rewritten as
\begin{align*}
I_{n,0}(p) &= \frac{1}{(n-1)!}\int_0^{\infty} dt F_n(\Omega(t))\omega(t)
\Bigg[  \Omega(t)^{n-1} h(\Omega(t))^2\\
& +
(n-1) \Omega(t)^{n-2} \left( 2h(\Omega(t))\int_0^t dt' \dot\omega(t') + \int_0^t dt'\frac{\dot \omega(t')^2}{\omega(t')}\right)\\
&+ (n-1)(n-2) \Omega(t)^{n-3}\left(\int_0^t dt'\dot \omega(t')\right)^2 \Bigg].
\end{align*}

We now consider the no-event part. First note that the total detection probability in case of $\Omega(\infty)<\infty$ was obtained in the main text as
\begin{align}\label{noeventp}
p_n^{\rm tot} = 1-\sum_{k=0}^{n-1}\frac{F_k(\Omega(\infty))}{k!} \Omega(\infty)^k.
\end{align}
It follows that the derivative has a simple expression, provided that $\dot \Omega(\infty)$ exists:
\begin{align}
\frac{d}{dp}p_n^{\rm tot} &= \dot \Omega(\infty) \sum_{k=0}^{n-1}\frac{F_{k+1}(\Omega(\infty))}{k!} \Omega(\infty)^k-\dot \Omega(\infty) \sum_{k=1}^{n-1}\frac{F_{k}(\Omega(\infty))}{(k-1)!} \Omega(\infty)^{k-1}\nonumber\\
&=\dot \Omega(\infty) \sum_{k=1}^{n}\frac{F_{k}(\Omega(\infty))}{(k-1)!} \Omega(\infty)^{k-1}-\dot \Omega(\infty) \sum_{k=1}^{n-1}\frac{F_{k}(\Omega(\infty))}{(k-1)!} \Omega(\infty)^{k-1} = \dot \Omega(\infty) \frac{F_{n}(\Omega(\infty))}{(n-1)!} \Omega(\infty)^{n-1}\label{noeventd}.
\end{align}
Combining this with the above formula for $I_{n,0}(p)$ gives the formula \eqref{fisher1} in the main text.

For the purpose of the time-stationary case and the further limits considered below, we further rewrite $I_{n,0}$ as 
\begin{align*}
I_{n,0}(p) &= \frac{1}{(n-1)!}\int_0^{\infty} dt F_n(\Omega(t))\omega(t) \Omega(t)^{n-1}\left[h(\Omega(t))^2 + (n-1)\left( 2h(\Omega(t))\frac{\dot\Omega(t)}{\Omega(t)} + \frac{\dot{\tilde\Omega}(t)}{\Omega(t)}+ (n-2) \left[\frac{\dot\Omega(t)}{\Omega(t)}\right]^2\right)\right],
\end{align*}
where
\begin{align*}
\dot\Omega(t) &= \int_0^t dt'\dot\omega(t'), & \dot{\tilde\Omega}(t)=\int_0^t dt'\frac{\dot \omega(t')^2}{\omega(t')}.
\end{align*}
Therefore, the Fisher information is given by
\begin{align*}
I_{n,0}&= \frac{1}{(n-1)!}\int_0^{\infty} dt F_n(\Omega(t))\omega(t) \Omega(t)^{n-1} S_n(\Omega(t)),
\end{align*}
where
\begin{align*}
S_n(\Omega(t)) &= h(\Omega(t))^2 + (n-1)\left( 2h(\Omega(t))\frac{\dot\Omega(t)}{\Omega(t)} + \frac{\dot{\tilde\Omega}(t)}{\Omega(t)}+ (n-2) \left[\frac{\dot\Omega(t)}{\Omega(t)}\right]^2\right), & h(\Omega(t)) &=-\frac{\Omega(t) \,F_{n+1}(\Omega(t))}{F_{n}(\Omega(t))}\frac{\dot \Omega(t)}{\Omega(t)} + \frac{\dot\omega(t)}{\omega(t)}.
\end{align*}
We can further simplify the function $S_n$; we denote $H_n(u) = F_{n+1}(u)/F_n(u)$, and obtain
\begin{align*}
S_n(\Omega(t)) &= \left[h(\Omega(t)) + (n-1)\frac{\dot\Omega(t)}{\Omega(t)}\right]^2- (n-1)\left[\frac{\dot\Omega(t)}{\Omega(t)}\right]^2+ (n-1) \frac{\dot{\tilde\Omega}(t)}{\Omega(t)}\\
&= \left[-\Omega(t) \,H_{n}(\Omega(t))\frac{\dot \Omega(t)}{\Omega(t)} + \frac{\dot\omega(t)}{\omega(t)} + (n-1)\frac{\dot\Omega(t)}{\Omega(t)}\right]^2- (n-1)\left[\frac{\dot\Omega(t)}{\Omega(t)}\right]^2+ (n-1) \frac{\dot{\tilde\Omega}(t)}{\Omega(t)}\\
&= \left[[n-1-\Omega(t) \,H_{n}(\Omega(t))]\frac{\dot \Omega(t)}{\Omega(t)} + \frac{\dot\omega(t)}{\omega(t)} \right]^2 + (n-1)\left[\frac{\dot{\tilde\Omega}(t)}{\Omega(t)}-\left[\frac{\dot\Omega(t)}{\Omega(t)}\right]^2\right].
\end{align*}
To summarise (for the purpose of the next subsections), 
\begin{align*}
I_{n,0}(p)&= \frac{1}{(n-1)!}\int_0^{\infty} dt F_n(\Omega(t))\omega(t) \Omega(t)^{n-1} \left(\left[[n-1-\Omega(t) \,H_{n}(\Omega(t))]\frac{\dot \Omega(t)}{\Omega(t)} + \frac{\dot\omega(t)}{\omega(t)} \right]^2 + (n-1)\left[\frac{\dot{\tilde\Omega}(t)}{\Omega(t)}-\left[\frac{\dot\Omega(t)}{\Omega(t)}\right]^2\right]\right).
\end{align*}

\subsubsection{The time-stationary case}
Consider now the (hypothetical) case where $\omega(t) = \omega_0$ (constant), which clearly has $\Omega(\infty) =\infty$. This case does not directly arise in our model because of the initial effects of the detection process, but we will show below that the Fisher information for limit of sparse beam (small spatial particle density) will coincide with the one obtained for the this case. In order to calculate it we note that $\Omega(t) = t\omega_0$, $\dot \Omega(t) =\dot \omega_0 t$, $\dot{\tilde \Omega}(t) = \frac{\dot\omega_0^2}{\omega_0} t$. The Fisher information becomes
\begin{align*}
I_n(p)&= \frac{1}{(n-1)!}\int_0^\infty dt F_n(\omega_0t)\omega_0 (\omega_0t)^{n-1} \left(\left[[n-1-\omega_0 t \,H_{n}(\omega_0 t)]\frac{\dot \omega_0}{\omega_0} + \frac{\dot\omega_0}{\omega_0} \right]^2 + (n-1)\left[\frac{\dot{\omega}_0^2}{\omega_0^2}-\left[\frac{\dot\omega_0}{\omega_0}\right]^2\right]\right)= C^{(n)}\left[\frac{\dot{\omega}_0}{\omega_0}\right]^2,
\end{align*}
where we have substituted $u = \omega_0 t$, and denoted
\begin{align}\label{stationaryconstant}
C^{(n)}=\frac{1}{(n-1)!}\int_0^\infty du F_n(u) u^{n-1} [n-u \,H_{n}(u)]^2.
\end{align}
It remains to evaluate the constant in the two cases. In the coherent case, we have $F_n(u) = e^{-u}$ so $H_n(u) = 1$; we use $\int_0^\infty du \,u^k e^{-u} = k!$ to get 
\begin{align*}
C^{(n)}&=\frac{1}{(n-1)!}\int_0^\infty du\, e^{-u} u^{n-1} [n-u]^2\\
&=\frac{1}{(n-1)!}\int_0^\infty du\, e^{-u} (n^2u^{n-1}-2nu^n+u^{n+1})\\
&=\frac{1}{(n-1)!}(n^2(n-1)!-2n n!+(n+1)!)\\
&=\frac{1}{(n-1)!}(n^2(n-1)!-2n^2 (n-1)!+(n+1)n(n-1)!)\\
&=n^2-2n^2 +(n+1)n =n.
\end{align*}
In the quasi-free case we have $F_n(u) = n!/(u+1)^{n+1}$ so $H_n(u) = (n+1)/(u+1)$; we use $\int_0^\infty du\, u^{k-1}/(1+u)^{k+1} =1/k$ to get 
\begin{align*}
C^{(n)}&=\frac{1}{(n-1)!}\int_0^\infty du\, \frac{n! u^{n-1}}{(1+u)^{n+1}} \left[n-(n+1)\frac{u}{1+u}\right]^2\\
&=n\int_0^\infty du\, \frac{u^{n-1}}{(1+u)^{n+1}} \left[n^2-\frac{2n(n+1)u}{1+u}+(n+1)^2\frac{u^2}{(1+u)^2}\right]\\
&=n\int_0^\infty du\, \left[n^2\frac{u^{n-1}}{(1+u)^{n+1}}-2n(n+1)\frac{u^{n}}{(1+u)^{n+2}} +(n+1)^2 \frac{u^{n+1}}{(1+u)^{n+3}}\right]\\
&=n\left[\frac{n^2}{n}-\frac{2n(n+1)}{n+1} + \frac{(n+1)^2}{n+2}\right]\\
&=n\left[\frac{n^2(n+1)(n+2) -2n^2(n+1)(n+2)+n(n+1)^3}{n(n+1)(n+2)}\right]\\
&=\left[\frac{n^2(n+2) -2n^2(n+2)+n(n+1)^2}{n+2}\right]\\
&=\left[\frac{-n^2(n+2)+n(n+1)^2}{n+2}\right]=\frac{n}{n+2}.
\end{align*}
Therefore, the Fisher information for the time-stationary case is
\begin{align*}
I_{n,0}(p) &= n \left(\frac{\dot \omega_0}{\omega_0}\right)^2 \text{  (coherent)}, & I_{n,0}(p) & = \frac{n}{n+2} \left(\frac{\dot \omega_0}{\omega_0}\right)^2 \text{  (quasi-free)}.
\end{align*}

\subsubsection{The limits $r_0\rightarrow 0$ and $r_0\rightarrow \infty$ in the case of uniform particle beam}

We now focus on the beam case considered in the main text, so that $\Omega(\infty) = \infty$. We consider the limit of particle density going to zero and infinity, so we write $\omega(t) = r_0g(t)$ and $G(t) =\int_0^t dt' g(t')$, where $g(t)=a|T_p + \tilde R_p(t)|^2$ is given by \eqref{delta_intensity}. It then also follows that $\dot \omega(t) = r_0 \dot g(t)$, where the derivative can be computed explicitly. \emph{We use (only) the existence of the limits}
\begin{align*}
g_\infty &:= \lim_{t\rightarrow \infty} g(t), & \dot g_\infty &:= \lim_{t\rightarrow \infty} \dot g(t),\\
g_0 &:= \lim_{t\rightarrow 0} g(t), & \dot g_0 &:= \lim_{t\rightarrow \infty} \dot g(t),
\end{align*}
and $g_0, g_\infty\neq 0$. These follow from the asymptotic relations for $\omega(t)$ and its derivative: \eqref{omegaapprox}, \eqref{domegaapprox} (for large $t$) and \eqref{omegaapproxsmall}, \eqref{domegaapproxsmall} (for small $t$), and the continuity of $g(t)$ and $\dot g(t)$ at $t=0$. \emph{In particular, $g_0$ and $g_\infty$ are both nonzero}; this is crucial in the following calculations.

Consider the following functions of $t$ appearing in the expression of the Fisher information:
\begin{align*}
\Phi(t)&:= \frac{\dot \Omega(t)}{\Omega(t)}=\frac{\int_0^t dt'\dot g(t')}{G(t)}, & \tilde\Phi(t)&:=  \frac{\dot{\tilde\Omega}(t)}{\Omega(t)}=\frac{1}{G(t)}\int_0^t dt'\frac{\dot g(t')^2}{g(t')}, & \phi(t) &:= \frac{\dot \omega(t)}{\omega(t)}=\frac{\dot g(t)}{g(t)}.
\end{align*}
The key point is that these are \emph{independent of $r_0$}. Now we rewrite the Fisher information of $p=p_0$ using the change of variables $u =\Omega(t)$ introduced earlier, but now taking into account that $\Omega(t) = r_0G(t)$, so that $t=G^{-1}(\tfrac{u}{r_0})$. The Fisher information is
\begin{align}\label{fisherr}
I_{n}^{(r_0)}(p_0)&= \frac{1}{(n-1)!}\int_0^{\infty} du F_n(u) u^{n-1} S_{n,r_0}(u),
\end{align}
where now \emph{all $r_0$-dependence in the integrand is in the function $S_{n,r_0}(u)$, and enters through $G^{-1}(\tfrac{u}{r_0})$ only}; explicitly,
\begin{align}\label{snr}
S_{n,r_0}(u) &=\left[(n-1-u \,H_{n}(u))\Phi(G^{-1}(\tfrac {u}{r_0})) + \phi(G^{-1}(\tfrac{u}{r_0})) \right]^2 + (n-1)\left[\tilde \Phi(G^{-1}(\tfrac{u}{r_0}))-\Phi(G^{-1}(\tfrac{u}{r_0}))^2\right].
\end{align}

We now consider the limits of the three functions $\Phi(t),\tilde\Phi(t),\phi(t)$: it follows directly by l'Hopital's rule that
\begin{align}
\lim_{t\rightarrow \infty} \Phi(t) &= \lim_{t\rightarrow \infty} \frac{\dot g(t)}{g(t)} =\lim_{t\rightarrow\infty}\phi(t)= \frac{\dot g_\infty}{g_\infty}, & \lim_{t\rightarrow \infty} \tilde\Phi(t) &=\lim_{t\rightarrow \infty} \frac{1}{g(t)}\frac{\dot g(t)^2}{g(t)} = \left[\frac{\dot g_\infty}{g_\infty}\right]^2,\label{limits}\\
\lim_{t\rightarrow 0} \Phi(t) &= \lim_{t\rightarrow 0} \frac{\dot g(t)}{g(t)} =\lim_{t\rightarrow 0}\phi(t)= \frac{\dot g_0}{g_0}, & \lim_{t\rightarrow 0} \tilde\Phi(t) &=\lim_{t\rightarrow 0} \frac{1}{g(t)}\frac{\dot g(t)^2}{g(t)} = \left[\frac{\dot g_0}{g_0}\right]^2,\label{limits2}
\end{align}
where each of the limits is now finite by our assumption. In particular, this implies that the functions $\Phi(t)$, $\tilde\Phi(t)$, and $\phi(t)$ (being continuous) are bounded on the whole interval $[0,\infty)$, and since they do not depend on $r_0$, there exist a constant $A>0$ only depending on $p$ (not $r_0$ or $t$) such that
\begin{align*}
\left|\Phi(t)\right|&\leq A, & |\tilde \Phi(t) |&\leq A, &\left|\phi(t)\right|&\leq A, & \text{for all }t>0.
\end{align*}
It follows that
\begin{align*}
|S_{n,r_0}(u)|&\leq A^2 \left[n+u \,H_{n}(u)\right]^2 + (n-1)A(1+A)\quad  \text{for all } u>0,
\end{align*}
and hence
\begin{align*}
|F_n(u) u^{n-1} S_{n,r_0}(u)| \leq F_n(u) u^{n-1}\left (A^2 \left[n+u \,H_{n}(u)\right]^2 + (n-1)A(1+A)\right),\quad \text{ for all }u>0,
\end{align*}
where the ($r_0$-independent) function on the right hand side turns out to be integrable. To prove this integrability rigorously, we need to consider separately the coherent and the quasi-free cases. In the coherent case, $H_n(u)=1$ and $F_n(u) = e^{-u}$, so the integral is
\begin{align}\label{boundintegral1}
\int_0^{\infty} du\, e^{-u} u^{n-1} \left(A^2 \left[n+u \right]^2 + (n-1)A(1+A)\right) <\infty.
\end{align}
In the quasi-free case we instead have $F_n(u) = n!/(u+1)^{n+1}$ and $H_n(u) = (n+1)(1+u)^{-1}$ so $u H_n(u) \leq n+1$ for all $u$, and hence the integral is bounded above by
\begin{align}\label{boundintegral2}
\int_0^{\infty} du\,\frac{n! u^{n-1}}{(1+u)^{n+1}} \left(A^2 \left[2n+1\right]^2 + (n-1)A(1+A)\right) <\infty.
\end{align}

The finiteness of the integrals \eqref{boundintegral1} and \eqref{boundintegral2} implies by dominated convergence theorem that we can take the limits $r_0\rightarrow 0$ and $r_0\rightarrow\infty$ in the Fisher information \eqref{fisherr} pointwise (for each fixed $u>0$) under the integral sign. Noting that $$\lim_{r_0\rightarrow 0}G^{-1}(\tfrac{u}{r_0}) = \infty\quad \text{for each }u>0$$ (because $G(\infty) = \infty$ and $G$ is invertible), we can apply the limits \eqref{limits} to \eqref{snr} to get
\begin{align*}
\lim_{r_0\rightarrow 0} S_{n,r_0}(u) &= \left[[n-1-u \,H_{n}(u)]\frac{\dot g_\infty}{g_\infty} + \frac{\dot g_\infty}{g_\infty} \right]^2 + (n-1)\left[\left[\frac{\dot g_\infty}{g_\infty}\right]^2-\left[\frac{\dot g_\infty}{g_\infty}\right]^2\right]\\
&= [n-u \,H_{n}(u)]^2\left[\frac{\dot g_\infty}{g_\infty}\right]^2, \quad \text{ for each }u>0,
\end{align*}
and a similar argument establishes the limit $r_0\rightarrow\infty$. Therefore, the Fisher information converges as
\begin{align*}
\lim_{r_0\rightarrow 0} I_n^{(r_0)}(p_0)&= C^{(n)} \, \left[\frac{\dot g_\infty}{g_\infty}\right]^2, &
\lim_{r\rightarrow \infty} I_n^{(r_0)}(p_0)&= C^{(n)} \, \left[\frac{\dot g_0}{g_0}\right]^2,
\end{align*}
where the constant $C^{(n)}$ is given by \eqref{stationaryconstant}. Interestingly, $\dot g_\infty \neq 0$ while $\dot g_0=0$ (see main text), so noting also that $g_0>0$ and $g_\infty >0$, we obtain:
\begin{align*}
\lim_{r_0\rightarrow 0}I_n^{(r_0)}(p_0) &= n \left(\frac{\dot g_\infty}{g_\infty}\right)^2 \text{  (coherent)}, & \lim_{r_0\rightarrow 0}I_n^{(r_0)}(p_0) & = \frac{n}{n+2} \left(\frac{\dot g_\infty}{g_\infty}\right)^2 \text{  (quasi-free)},\\
\lim_{r_0\rightarrow \infty}I_n^{(r_0)} &= 0 \text{  (coherent)}, & \lim_{r_0\rightarrow \infty}I_n^{(r_0)}(p_0) & =0\text{  (quasi-free)}.
\end{align*}

\end{widetext}


\begin{thebibliography}{99}

\bibitem{Muga2000}
J. G. Muga, C. R. Leavens, ``Arrival Time in Quantum Mechanics", Phys. Rep. {\bf 338}, 353 (2000).



\bibitem {Muga2002}%
  J. G. Muga, R. Sala Mayato and I. L. Egusquiza (eds.) {\it Time in quantum
    mechanics}
(Lect. Notes Phys. m72, Springer, Berlin) (2002).

\bibitem {Muga2009}%
  J. G. Muga, A. Ruschhaupt and A. del Campo (eds.) {\it Time in quantum
    mechanics -- Vol.2}
(Lect. Notes Phys. 789, Springer) (2009).


\bibitem{Kiukas2009} J. Kiukas, A. Ruschhaupt and R. F. Werner, ``Tunneling Times with Covariant Measurements", Found. Phys. {\bf 39}, 829 (2009).

\bibitem{Kiukas2013} J. Kiukas, A. Ruschhaupt and R. F. Werner, ``Full counting statistics of stationary particle beams'', J. Math. Phys. {\bf 54}, 042109 (2013).

\bibitem{Maccone2020} L. Maccone and K. Sacha, ``Quantum Measurement of Time", Phys. Rev. Lett. {\bf 124}, 110402 (2020).

\bibitem{Roncallo2023} S. Roncallo, K. Sacha and L. Maccone, ``When does a particle arrive?", Quantum {\bf 7}, 968 (2023).

\bibitem{Kijowski2023} J. Kijowski, ``Arrival Time in Quantum Mechanics", Acta Phys. Pol A {\bf 143}, s123 (2023).

\bibitem{Page1983}
D. N. Page and William K. Wootters, ``Evolution without evolution:
Dynamics described by stationary observables",
Phys. Rev. D {\bf 27}, 2885–2892 (1983).

\bibitem{Giovannetti2015}
V. Giovannetti, S. Lloyd and L. Maccone, ``Quantum Time", Phys. Rev. D {\bf 92}, 045033 (2015).

\bibitem{Marletto2017}
C. Marletto and V. Vedral, ``Evolution without evolution, and
without ambiguities”, Phys. Rev. D {\bf 95}, 043510 (2017).

\bibitem{Pauli1958} W. Pauli: in Encyclopedia of Physics, ed. by S. Flugge (Springer, Berlin), Vol. V/1, p. 60 305 (1958).


\bibitem{Aharonov1961}
Y. Aharonov and D. Bohm,
``Time in the Quantum Theory and the
Uncertainty Relation for Time and Energy'',
Phys. Rev. {\bf 122}, 1649 (1961).

\bibitem{Kijowski1974}
J.~Kijowski, ``On the Time Operator in Quantum Mechanics and the Heisenberg Uncertainty Relation for Energy and Time",
Rep. Math. Phys. {\bf 6}, 361 (1974).

\bibitem {Muga2002a}%
  I. L. Egusquiza, J. G. Muga and A. D. Baute, ``{`Standard'} quantum mechanical
  approach to times of arrival'', in \cite{Muga2002}

\bibitem{Werner1986} R. Werner, ``Screen observables in relativistic and nonrelativistic quantum mechanics", J. Math. Phys. {\bf 27}, 793 (1986).

\bibitem {Werner1989}%
  R.~F. Werner, ``Inequalities expressing the {P}auli
  principle for generalized observables", in Mathematical methods in statistical mechanics (Leuven Univ. Press), pp. 179 -196 (1989).

\bibitem{Baute2002} 
A.D. Baute, I.L. Egusquiza, J.G. Muga, ``Quantum times of arrival for multiparticle states", Phys. Rev. A 
{\bf 65}, 032114 (2002).



\bibitem{Damborenea2002}
J.A.~Damborenea, I.L.~Egusquiza, G.C.~Hegerfeldt, J.G.~Muga, ``Measurement-based approach to quantum arrival times",
Phys. Rev. A {\bf 66}, 052104 (2002).

\bibitem {Ruschhaupt2009}%
  A. Ruschhaupt, J. G. Muga and G. C. Hegerfeldt, ``Detector models
  for the quantum time of arrival'', in \cite{Muga2009}


\bibitem{Allcock1969}
G.R. Allcock, ``The Time of Arrival in Quantum Mechanics", Ann. Phys. (N.Y.) {\bf 53}, 253 (1969); {\bf 53}, 286
(1969); {\bf 53}, 311 (1969).

\bibitem{Werner1987} R. Werner, ``Arrival Time Observables in Quantum Mechanics", Ann. Inst. Henri Poincar\'e {\bf 47}, 429 (1987).

\bibitem{Ruschhaupt2004} A. Ruschhaupt, J.A. Damborenea, B. Navarro, J.G. Muga, 
G.C. Hegerfeldt,
``Exact and approximate complex potentials for modelling time observables", 
Europhys. Lett. {\bf  67}, 1 (2004).

\bibitem{Ruschhaupt2004a}
A. Ruschhaupt, B. Navarro, J. G. Muga, ``Perfect detection of ultra-cold atoms by laser-induced
ionization", J. Phys. B: At. Mol. Opt. Phys. {\bf 37}, L313 (2004).
                  
\bibitem{Kiukas2012} J. Kiukas, A. Ruschhaupt, P. O. Schmidt and R. F. Werner, ``Exact energy–time uncertainty relation for arrival
time by absorption", J. Phys. A: Math. Theor. {\bf 45}, 185301 (2012).



\bibitem{Hegerfeldt1992}  G.C. Hegerfeldt, T.S. Wilser, in
{\it Classical and Quantum Systems.} 
Proceedings of the Second International Wigner Symposium, July
1991, ed. by H.D. Doebner, W. Scherer, F. Schroeck, (World
Scientific, Singapore, 1992), p. 104;
G.C. Hegerfeldt,
Phys. Rev. A {\bf 47}, 449 (1993); G.C. Hegerfeldt, 
D.G. Sondermann, Quantum 
  Semiclass.~Opt.~{\bf 8}, 121 (1996). For a review cf.  M.B. Plenio, P.L. Knight, 
Rev. Mod. Phys. {\bf 70}, 101 (1998).

\bibitem{Dalibard1992}
{J. Dalibard, Y. Castin, K.  M{\o}lmer},
{Phys. Rev. Lett.} {\bf 68}, {580} (1992)

\bibitem{Carmichael1993} 
H. Carmichael,  {\it 
An Open Systems Approach to Quantum
Optics}
(Springer, Berlin, 1993).


\bibitem{Davies1976} E. P. Davies, {\it Quantum Theory of Open Systems}, Academic Press, London 1976.

\bibitem{Barchielli1986} A. Barchielli, ``Measurement theory and stochastic differential equations in quantum mechanics", Phys. Rev. A. {\bf 34} 1642 (1986).

\bibitem{Fannes1992} M. Fannes, B. Nachtergaele, R.F. Werner, ``Finitely correlated states on quantum spin chains", Commun. Math. Phys. {\bf 144} 443 (1992).

\bibitem{Verstraete2010} F. Verstraete and J. I. Cirac, ``Continuous matrix product states for quantum fields", Phys. Rev. Lett. {\bf 104} 190405 (2010).

\bibitem{Ciccarello2022} F. Ciccarello, S. Lorenzo, V. Giovannetti, G. M. Palma,
``Quantum collision models: Open system dynamics from repeated interactions",
Physics Reports {\bf 954} 1-70 (2022).


\bibitem{Lindblad1976} G. Lindblad, Commun. Math. Phys. {\bf 48}, 119-130 (1976).

\bibitem{Gorini1976} V. Gorini, A. Kossakowski, E. C. G. Sudarshan, Math. Phys. {\bf 17}, 821 (1976).


\bibitem {Alicki2007} R. Alicki, ``General Theory and Applications to Unstable Particles", Lect. Notes Phys. 717, 1-46, Springer, Berlin (2007)

\bibitem {Butz2010} M. Butz and H. Spohn, ``Dynamical phase transition
  for a quantum particle source", Ann. Henri Poincar\'e {\bf 10}, 1223--1249 (2010).
 
\bibitem {Benard1973}%
 C. B\'enard and O. Macchi, ``Detection and
 'emission' processes of quantum particles in a 'chaotic state'", J. Math. Phys. {\bf 14}, 155 - 167 (1973).



\bibitem{Bratteli1997}  O. Bratteli and D. W. Robinson, ``Operator Algebras and Quantum Statistical Mechanics II" (Springer, Berlin, 1997).

\bibitem{Daley1988} D. Daley and D. Vere-Jones, ``An Introduction to the Theory of Point Processes" (2 vols.) (Springer, 1988).

\bibitem{Snyder1991} D. L. Snyder and M. I. Miller, ``Random point processes in time and space" (Springer New York, 1991). 

\bibitem{Gregoire1983} G. Gregoire,  ``Negative binomial distributions for point processes, Stochastic processes and their applications {\bf 16} 179 (1983).

\bibitem{Abramovitz1965} M. Abramowitz, I. A. Stegun, ``Handbook of Mathematical Functions" (Dover Publications, Inc., New York, 1965)

\bibitem{Lew1972} J. S. Lew, ``On Linear Volterra Integral Equations of Convolution Type'', Proceedings of the American Mathematical Society. {\bf 35}, 450-456 (1972).

\bibitem{Bellman1984} R. E. Bellman, R. S. Roth, ``The Laplace transform", World Scientific, Singapore, 1984.



\bibitem{Garthwaite2006} P. H. Garthwaite, I. T. Jolliffe, B. Jones, ``Statistical Inference", 2nd Edition (Oxford University Press, New York, 2006).

\bibitem{Paris2009} M. A. Paris, ``Quantum estimation for quantum technology", Int. J. Quant. Inf. {\bf 7} 125 (2009).

\end{thebibliography}
\end{document}